\documentclass[10pt,a4paper,twocolumn]{article}
\usepackage{times}
\usepackage[top=20mm,left=24mm,right=24mm,bottom=30mm,columnsep=5mm]{geometry}

\usepackage[small,compact]{titlesec}

\usepackage[utf8]{inputenc}
\usepackage[T1]{fontenc}
\usepackage{graphicx}
\usepackage{longtable}
\usepackage{float}
\usepackage{wrapfig}
\usepackage{rotating}
\usepackage[normalem]{ulem}
\usepackage{amsmath}
\usepackage{textcomp}
\usepackage{marvosym}
\usepackage{wasysym}
\usepackage{amssymb}
\usepackage[table,usenames,dvipsnames]{xcolor}
\usepackage{verbatim}
\usepackage{subcaption}

\usepackage{dblfloatfix} % https://stackoverflow.com/questions/53167913/how-to-place-the-figure-at-the-bottom-of-page-in-latex

\usepackage{gensymb} % celsius

\usepackage{hyperref}

\usepackage{tocbibind}
\tocotherhead{section}

\usepackage[labelfont=bf]{caption} % my style

\let\origdoublepage\cleardoublepage
\newcommand{\clearemptydoublepage}{%
  \clearpage
  {\pagestyle{empty}\origdoublepage}%
}
\let\cleardoublepage\clearemptydoublepage

\usepackage{fancyhdr}

\fancypagestyle{plain}{%
\fancyhf{}
\fancyfoot[C]{\small{\thepage}}

}

\newcommand{\papertrc}{Paper~A}
\newcommand{\ptrc}{A}

\newcommand{\paperissre}{Paper~B}
\newcommand{\pissre}{B}

\newcommand{\papermapper}{Paper~C}
\newcommand{\pmapper}{C}

\newcommand{\paperflaky}{Paper~D}
\newcommand{\pflaky}{D}

\newcommand{\papertim}{Paper~E}
\newcommand{\ptim}{E}

\newcommand{\papertrdb}{Paper~X3}

\newcommand{\paperethics}{Paper~X5}
\newcommand{\pethics}{X5}

\newcommand{\pairts}{X6}

\usepackage{authblk}
\author[1,2]{Per Erik Strandberg}
\affil[1]{Westermo Network Technologies AB, Västerås, Sweden%
\authorcr Email: per.strandberg@westermo.com%
\authorcr \hspace{1.5in}}
\affil[2]{Mälardalen University, Västerås, Sweden%
\authorcr Email: per.erik.strandberg@mdh.se}

\title{Automated System-Level Software Testing of Industrial Networked Embedded Systems}
\date{2021}

\begin{document}

\maketitle

\begin{figure*}[t!]
\centering
\includegraphics[width=0.8\linewidth]{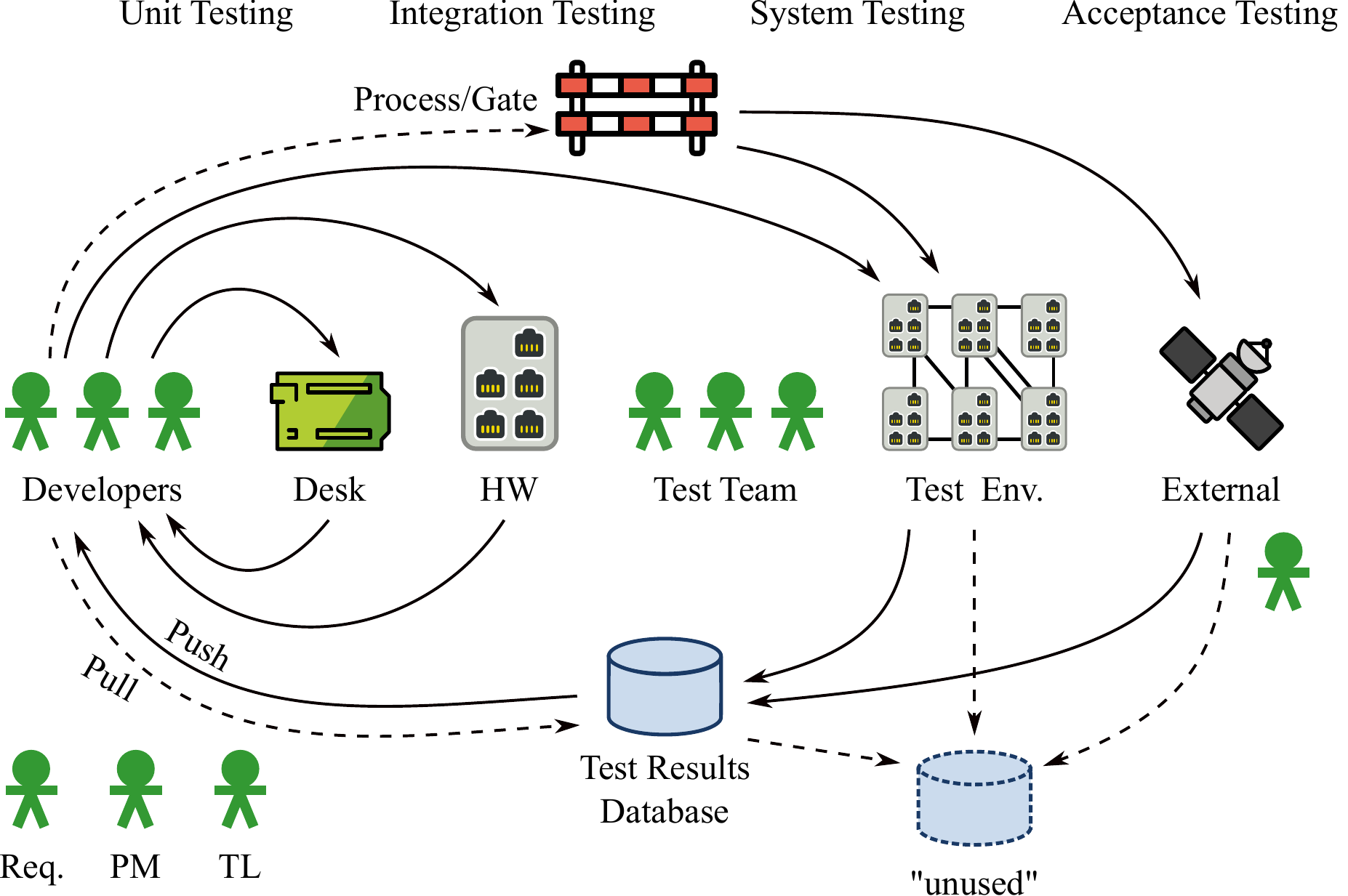}
\caption{
%Automated system-level software testing of industrial networked embedded systems.
In the software development of embedded systems, one has to test the software as it runs on physical devices.
The testing can be automated, and this thesis covers some of the challenges that can arrive: % in automated large-scale system-level software testing of industrial networked embedded systems:
How does the information related to testing flow?
How should test cases and hardware devices be selected for the testing?
Why do some test cases change verdict for no apparent reason?
Finally, how can test results be made explorable and visualized?
  % Flow of Information in Software Testing.
  \label{fig-contrib-flow}
  }
\end{figure*}

\noindent
\emph{And into the forest I go to lose my mind and find my soul.}

\noindent
\hspace{42mm}-- John Muir (1838-1914)

\section*{Abstract}

% Intro
%\remarkD{Did not focus on abstract this time.}
%\remarkP{Re-read this.}
Embedded systems are ubiquitous and play critical roles
in management systems for industry and transport.
Software failures in these domains
may lead to loss of production or even loss of life,
so the software in these systems needs to be reliable.
Software testing is a standard approach for quality assurance
of embedded software, and
many software development processes strive for test automation.
% Problem
%%% However, important challenges for successful software test automation are:
%%% lack of time for testing,
%%% lack of test environment availability,
%%% intermittently failing tests,
%%% as well as an excess of complex test results information.
%%% These challenges render decision-making in the software development process hard.
%%% \remarkP{
%%% Link between this paragraph and research goal.
%%% Talk more about system level?
%%% What is a good research goal?
%%% }
%%%
%
%%% \remarkP{The five problems}
% re-phrased 20200819 /Per
Out of the many challenges for successful software test automation,
this thesis addresses five:
% However, some major challenges
%
%are covered in this thesis:
%are:
(i)
understanding how updated software reaches a test environment, how testing is conducted in the test environment, and how test results reach the developers that updated the software in the first place;
%\remarkD{Phrased like this, isn't it just a technical context-specific problem? Discuss in mtg. Place last?};
(ii)
selecting which test cases to execute in a test suite given constraints on available time and test systems;
(iii)
given that the test cases are run on different configurations of connected devices,
selecting which hardware to use for each test case to be executed;
(iv)
analyzing test cases that, when executed over time on evolving software, testware or hardware revisions, appear to randomly fail;
and
(v)
making test results information actionable with test results exploration and visualization.

The challenges
%In this thesis these challenges
are tackled in several ways.
%% Method TRC + TRDB
First,
to better understand the flow of information in the embedded systems software development process, interviews at five different companies were conducted.
% Result TRC + TRDB
The results show how visualizations and a test results database support decision-making.
Results also describe the overall flow of information in software testing:
from developers to hardware in the test environment, and back to developers. %via the test results database.
%\remarkD{Should database really be here? Is this always (or even typically) the case?} % Method: lack of time
Second, in order to address the challenges of test selection and hardware selection,
automated approaches for testing %smarter \remarkD{use other word than "smarter"?} 
given resource constraints were implemented and evaluated using industrial data stemming from years of nightly testing.
% Result 1
It was shown that these approaches could solve problems such as nightly testing not finishing on time,
as well as increasing hardware coverage by varying hardware selection over test iterations.
% Method: Flaky
Third,
the challenge of intermittently failing tests was addressed with a new metric that can classify test cases as intermittently or consistently failing.
% Results flaky
Again, by using industry data,
factors that lead to intermittent failures were identified,
and similarities and differences between root causes
for intermittently and consistently failing tests were observed.
% TIM
Finally, in order to better render test results actionable, a tool was implemented for test results exploration and visualization.
The implementation was evaluated using a reference group and logging of the tool's usage. Solution patterns and views of the tool were identified, as well as challenges for implementing such a tool.

% Discussion
%System-level software testing of networked embedded systems
%can be difficult to automate. This thesis addresses several important
%challenges and provides results that are of interest both to
%industrial practitioners and researchers.

\section*{Note to Reader}
This is a compact version of the introduction (kappa) of my doctoral thesis. Some content has been 
removed or shortened, but the section numbers are kept the same.

\section*{List of Publications Included in this Thesis}

% experimental

\begin{description}
\item[\papertrc:] {P. E. Strandberg, E. P. Enoiu, W. Afzal, D. Sundmark, and R. Feldt.
``Information Flow in Software Testing -- An Interview Study with Embedded Software Engineering Practitioners.''
In \emph{IEEE Access,} 7:46434--46453, 2019
\cite{strandberg2019flow} (Instrument:~\cite{strandberg2018instrument}).

% I was the main driver of this study.

Presentation: \href{https://youtu.be/KVVVxe3dH8o}{https://youtu.be/KVVVxe3dH8o}
}

\item[\paperissre:]{P. E. Strandberg, D. Sundmark, W. Afzal, T. J. Ostrand, and E. J. Weyuker.
``Experience report: Automated System Level Regression Test Prioritization using Multiple Factors.''
In \emph{International Symposium on Software Reliability Engineering}, IEEE, 2016 \cite{strandberg2016}.

%I was the main driver of this study.
Winner of \textbf{best research paper} award at ISSRE'16.
}

\item[\papermapper:]{P. E. Strandberg, T. J. Ostrand, E. J. Weyuker, D. Sundmark, and W. Afzal.
``Automated Test Mapping and Coverage for Network Topologies.''
In \emph{International Symposium on Software Testing and Analysis}, ACM, 2018 \cite{strandberg2018automated}.

%I was the main driver of this study.
}

\item[\paperflaky:]{P. E. Strandberg, T. J. Ostrand, E. J. Weyuker, D. Sundmark, and W. Afzal.
``Intermittently Failing Tests in the Embedded Systems Domain.''
In \emph{International Symposium on Software Testing and Analysis}, ACM, 2020 \cite{strandberg2020intermittently}.

%I was the main driver of this study.

Presentation: \href{https://youtu.be/W1G5hVfp_Sw}{https://youtu.be/W1G5hVfp\_Sw}
}

\item[\papertim:]{P. E. Strandberg, W. Afzal and D. Sundmark.
``Software Test Results Exploration and Visualization with Continuous Integration and Nightly Testing.''

Submitted to Springer's International Journal on Software Tools for Technology Transfer (STTT) in July 2021.
}
\end{description}

%\setcounter{tocdepth}{1}
%\tableofcontents

% Not everyone likes the i, ii, iii, ... numbers in front matter.
% \mainmatter

\newcommand{\mychapter}{\section}
\newcommand{\mysection}{\subsection}
\newcommand{\mysubsection}{\subsubsection}

%\part{Thesis}
%\pagestyle{fancy}

%%%%%%%%%%%%%%%%%%%%%%%%%%%%%%%%%%%%%%%%%%%%%%%%%%%%%%%%%%%%%%%%%%%%%%%%%
\mychapter{Introduction}

This chapter starts with 
%In this section we introduce
a short history of industry-academia collaboration and industrial doctoral studies, before it introduces
the personal and industrial context of this thesis, 
and then it gives a background to the thesis.

%\remarkP{Idea: would the flow be better if this chapter \emph{started} with the brief history, and then had personal context, and then background?}

%%%%%%%%%%%%%%%%%%%%%%

%\remarkD{I like the idea of the below section, but I am not sure about where it should be. Here it is a detour.}
%\section{Brief History of Research in Industry-Academia Collaboration and Industrial Doctoral Studies}
\mysection[Brief History of Research in Industry-Academia\ldots]{Brief History of Research in Industry-Academia Collaboration and Industrial Doctoral Studies}

\begin{figure*}[t]
\centering
\includegraphics[width=0.8\linewidth]{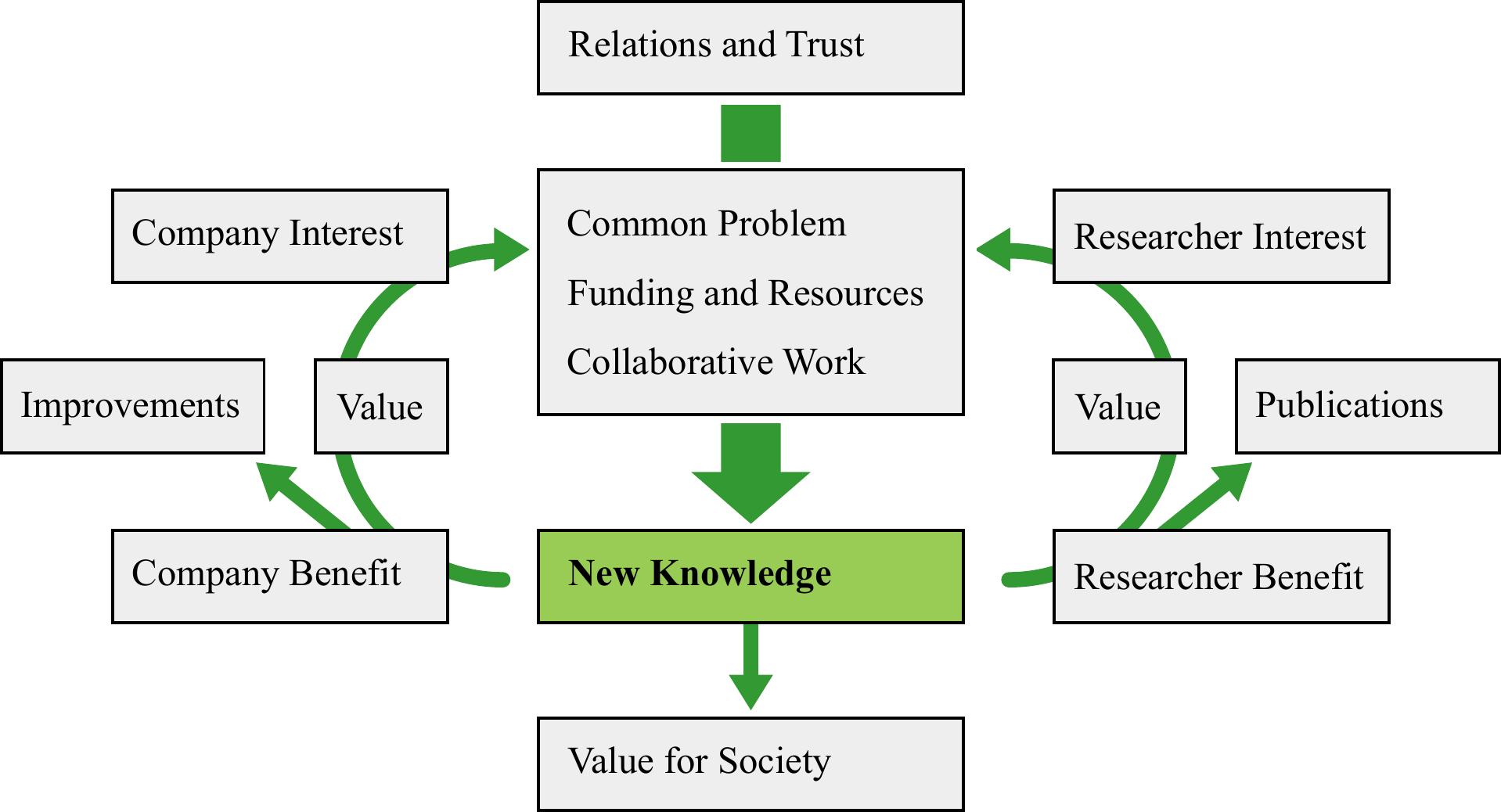}
\caption{
  Industry-Academia collaboration based on Sannö et al.~\cite{sanno2019increasing}.
\label{fig-collab}
%\remarkP{TODO: Use not as intense colors.}
}
\end{figure*}

\label{ind-acad-collab}

Arora et al.\ \cite{arora2020changing} investigate the changes in American innovation from 1850 and onwards.
In particular, they discuss the golden age of the corporate labs, with AT\&T's Bell Labs that employed 15 thousand people (roughly a tenth with PhDs) in the late 1960s. Fourteen Nobel Prizes and five Turing Awards were awarded to alumni at Bell Labs.
The authors mention that since the 1980s, the distances between universities and corporations have grown.
Universities focused on research on smaller and smaller problems, leading to innovation that could not be used at companies without great difficulty. Corporations instead focused on development.
At this time, corporate research started declining -- publications per firm decreased, and patents per firm has increased (except in the life sciences).
Some of the possible reasons for this decline is the difference in attitude towards research, where corporate research is often more mission-oriented, and university research sometimes curiosity-driven.
More concretely, the decline might be caused by (i) spill-over - where rivals use publications from a company in their patents, (ii) narrowed firm scope, (iii) increased distances between manufacturing and R\&D due to changes in trade, outsourcing, and offshoring. Finally, Arora et al.\ mention that (iv) tapping into knowledge and invention from external sources has been simplified.

According to Geschwind\ \cite{geschwind2018doctoral},
the Swedish doctoral training and degree was reformed in 1969, inspired by US systems.
Doctoral studies contain research and course work, corresponding to 4 years full time work.
In general, since the 1960s, there has been
an increased number of doctoral students,
and since the end of the last millennia also an increase in throughput.
The share of women has risen from 16\% 1962 to about 50\% in 2005.
There has also been a reduced admission in humanities
and about 40\% of admitted doctoral students in Sweden were recruited from abroad.
In 1977, a reform led to formation of several new universities in Sweden.
Three reasons for the reform were to provide universal access to higher education,
to limit outward migration stemming from industrial restructuring,
and to provide skilled employees to industry
\cite{assbring2017collab}.
According to Smith \cite{smith2017ethical},
in the mid 1990s
the Swedish foundation for the advancement of knowledge and competence
started offering part of funding for research if universities collaborated with industry.
By 2000, thirteen industrial research schools had started and
in 2017, Chalmers University of Technology alone had about 160 industrial doctoral students.

Assbring and Nuur \cite{assbring2017collab}\ made a case study on the perceived industrial benefits of participating in collaborative research school.
They found that motivators for companies include
access to new knowledge and state-of-the-art research,
competence creation and retention,
new or improved products and processes,
as well as legitimacy.
Small firms are more mission-oriented, %motivated of developing concrete products and services,
whereas larger firms are more motivated by more general goals such as developing competence.

\begin{figure*}[t]
\centering
\includegraphics[width=0.7\linewidth]{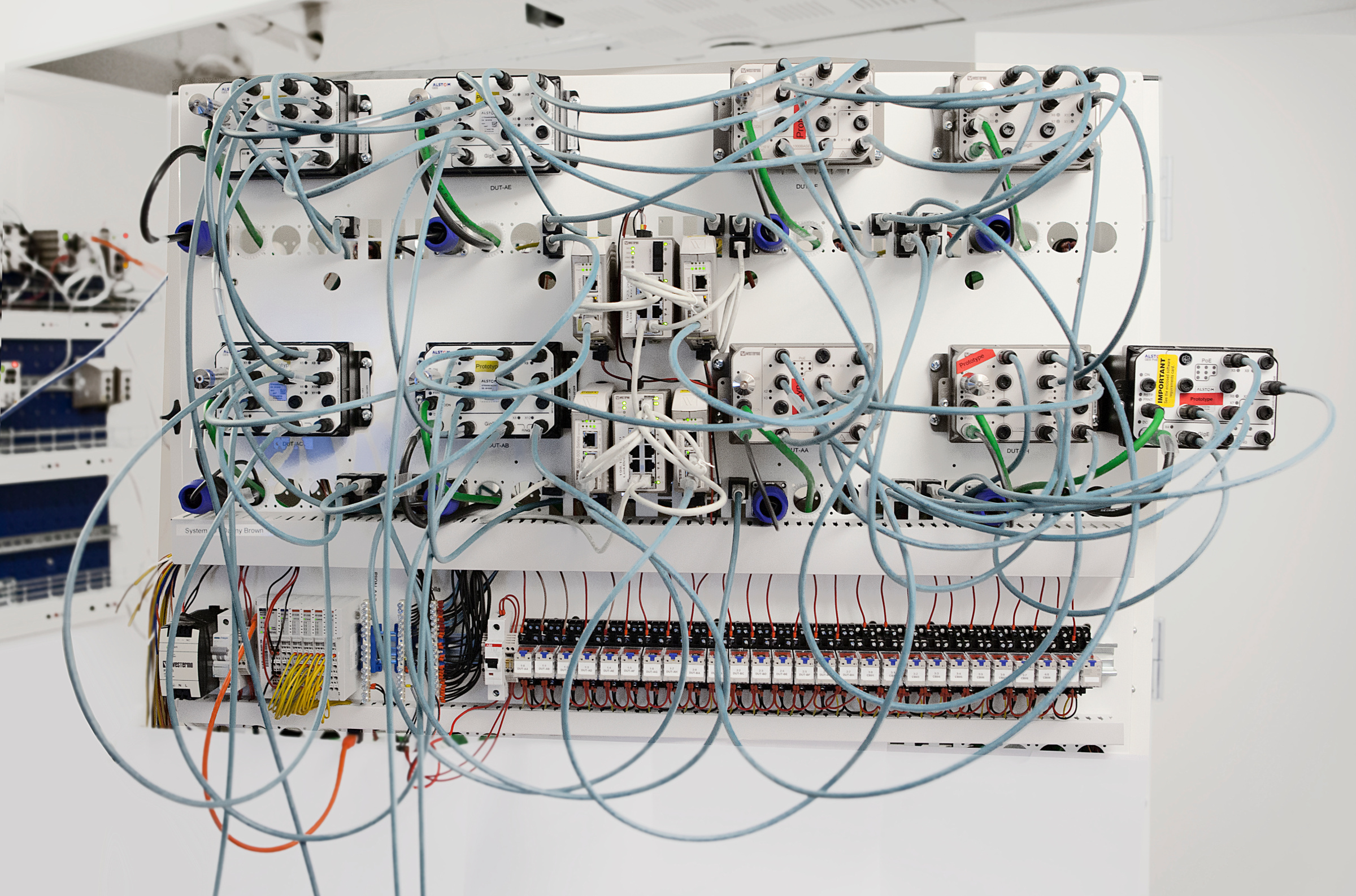}
\caption{
  An example of a Westermo test system.
  %A test system.
  \label{fig-kappa-test-system}
}
\end{figure*}

The Swedish Higher Education Act~\cite{swedish1992higher}
states that higher education institutions shall
involve external parties and ensure that they can benefit from research findings.
Doing co-production, i.e.\ having an industry-academia collaboration in research,
is not only a way to comply with regulations and recommendations, it can also be a suitable way to conduct relevant research.
%
%\remarkW{An introductory sentence on co-production required as it is not that commonly used term.}
%\remarkP{Above OK?}
For co-produced research to be \emph{relevant} and successful, the research party and the industry party should share a common understanding of the problem and be able to communicate~\cite{eldh2013some,garousi2016challenges,hove2005experiences,sanno2019increasing}.
This can be a challenge, as Sannö et al.\ point out~\cite{sanno2019increasing}, because these two parties typically have differences in perspective with respect to problem formulation, methodology, and result; as well as counterproductive differences in their views on knowledge and in their driving forces.
Lo et al.~\cite{lo2015practitioners} and Carver et al.~\cite{carver2016practitioners}
made studies in 2015 and 2016 on how practitioners perceived relevance of research in software engineering.
There seems to be no correlation between citation count and its perceived relevance,
and papers with industrial co-authors were only marginally more relevant.
% and practitioners welcomed research on code quality and sustainability/maintenance.
%
Sannö et al.\ discuss ways to increase the impact of industry-academia collaboration and present the model illustrated in Figure~\ref{fig-collab}.
As mentioned, the formulation of a common problem is central. The model also illustrates that the impact of the collaboration does not \emph{end} with new knowledge -- instead the new knowledge can be seen as what \emph{drives} further industry-academia research through different benefits.
Weyuker and Ostrand \cite{weyukerostrand2017experiences} recommend that academics strive to involve at least one industry participant that is committed to the goal of the study, and that the industry partner sees the research as relevant, valuable and important.
Eldh points out that an interest in continued partnership is a good evaluation criteria for both industry and academia~\cite{eldh2013some}.
A more detailed approach for industry-academia collaboration was proposed by Marijan and Gotlieb \cite{marijan2021industry}. Their process involves a model in seven phases starting with 
problem scoping (focusing on the industrial problem) and knowledge conception (to formulate research problem); followed by development, transfer, exploitation and adoption of knowledge and technology; and ending in market research in order to explore if benefits from the research could reach outside of the project.
All research problems and questions in this thesis have been formulated jointly while involving both industry and academia.

There is no doubt a great value in corporate research. Arora et al.\ \cite{arora2020changing} argue that these labs solve practical problems, three of the reasons are that companies have the ability to test innovation at scale by having large data sets, companies are often multi-disciplinary and may have unique equipment.
Furthermore, doctoral studies is no longer only perceived as a preparation for an academic career,
PhD holders are attractive to industry \cite{assbring2017collab}.
Finally, many industrial doctoral students feel privileged to carry out research in close collaboration with industry \cite{assbring2017collab}.

% 
% \begin{figure*}[p!]
% \centering
% \includegraphics[width=0.9\textheight,angle=90]{./images/time-line-2011-phd.pdf}
% \caption{
%   Time line from 2011 to 2022.
%   \remarkP{TODO: Remove}
%   \remarkP{Many comments from Wasif -- see email.}
%   \label{fig-timeline-2011-2021}
% }
% \end{figure*}
% 

\mysection{Personal and Industrial Context}

\begin{figure*}[t]
\centering
\includegraphics[width=0.6\linewidth]{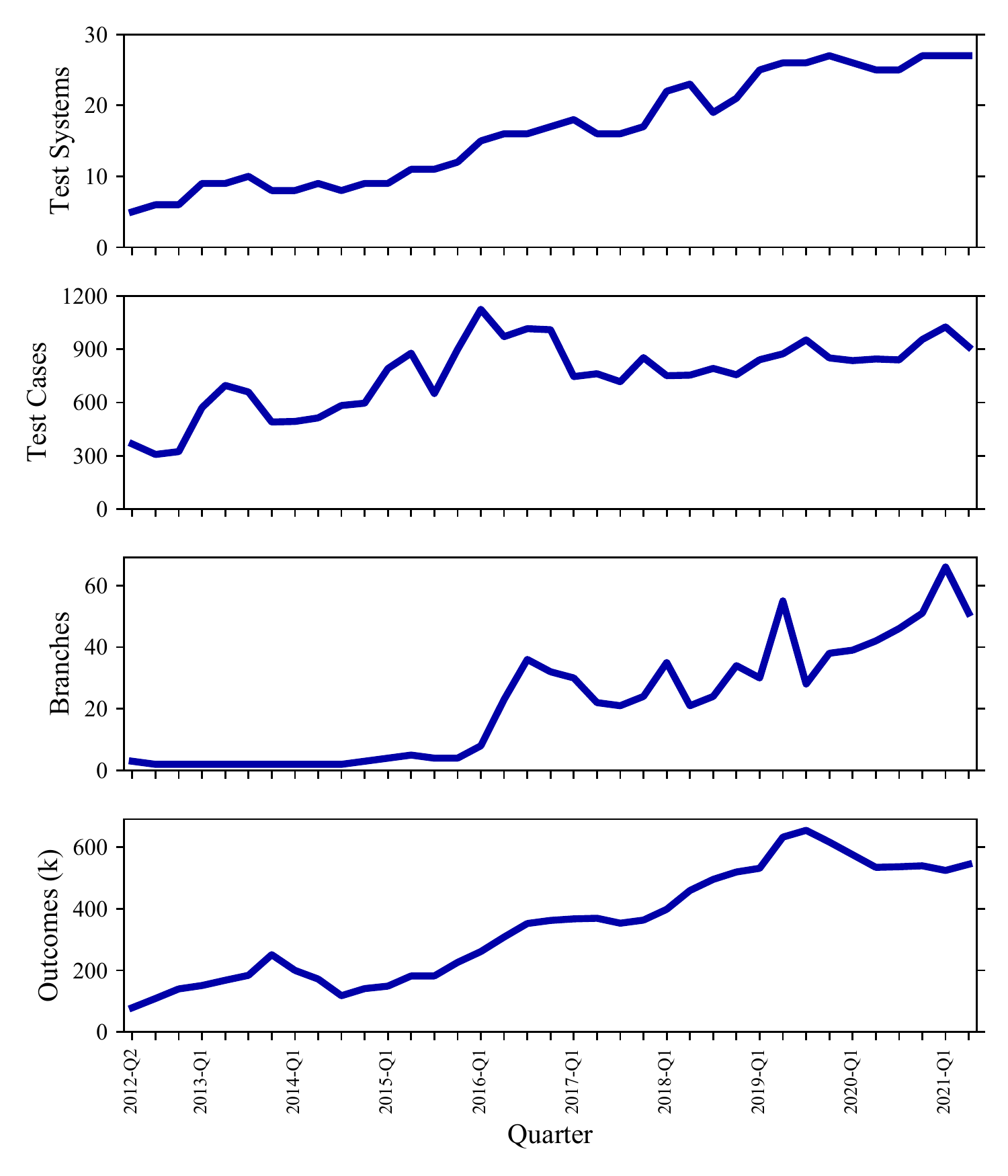}
\caption{
Growth in test complexity at Westermo between 2012 and Q2 2021.
% Growth in test complexity from 2012 to Q2 2021.
\label{20200312-growth-line}
}
\end{figure*}

%\remarkP{Cite Patton here?}

%\remarkP{Postpone acronym definitions -- not in this section.}

%\remarkP{Move here?}
Providing information on \emph{context} in software engineering research is relevant in order to allow readers to draw valid conclusions from
the research~\cite{dybaa2012works,pw-context-2009,runeson2012}.
In addition, in research dealing with qualitative data, the researcher is ``\emph{active} in the research process'' and
there is a risk for researcher bias where personal background and industrial context could play a role \cite{braun-clarke-2006,patton2014qualitative}. %\remarkD{I think you could easily find references to support this claim. It's (afaik) common practice in other fields.}
Here I mention context such that a reader could take
this into consideration when reading this thesis.
%(overview in Figure~\ref{fig-timeline-2011-2021}).

Before starting as an industrial doctoral student in 2017, I worked
for 11 years with software testing, development, and requirements in
the rail, nuclear, web, and communication equipment domains.  During
the majority of this time, I was a consultant, and as such, I led a
competence network on software testing for 5 years.
In this period of my life, I studied many test and requirements certification
syllabi and became a certified tester
as well as a certified professional for requirements engineering
(ISTQB foundation, ISTQB test manager, ISTQB agile tester,
and REQB CPRE Foundation).

I am employed full time at Westermo Network Technologies AB (Westermo),
where I have worked with test automation, test management and research.
%
%Westermo
The company designs and manufactures robust data communication devices
for harsh environments, providing communication infrastructure for
control and monitoring systems where consumer grade products are not
sufficiently resilient.
These devices and the software in them, %the Westermo Operating System,
are tested nightly with the main purpose of finding
software regressions.
In order to test for regressions, a number of test systems
%such as the ones in the rightmost parts of Figure~\ref{test-levels}
built up of physical or virtualized devices under test running the software under test have been constructed
(see example in \ref{fig-kappa-test-system}).
Automated test cases have also been implemented, as well
as a test framework for running the test cases.
I first added code to this framework in 2011, when it was
being migrated into what is now its current form.
% The test systems are
% build up of DUTs running WeOS, and the test
% framework communicates with the DUTs using a command line interface
% over serial port communication.
More recently, this framework has started a transition from Python into the Go
programming language.
Over time, more and more components
have been added to this eco-system for nightly testing,
such as the test selection process described in \paperissre{},
the test case mapping onto test systems described in \papermapper{},
and the test results database that is central to \papertim.
%, and the process for starting nightly testing has become complex.

In 2014, when the implementation of a test selection tool was evaluated, I wrote an internal report at Westermo. At this time, I was encouraged to improve the report and publish it in an academic context by the software manager. Around the same time, I attended a course for becoming ISTQB Certified Test Manager. Through luck, connections and curiosity I got in contact with the Software Testing Laboratory at MDH.
After some encouragement and coaching on how to write an academic paper, I submitted an early version of \paperissre\ without coauthors.
After being rejected with very constructive feedback, we conducted an evaluation of changes in fault distribution. Together with four other authors, I submitted to ISSRE in 2016 and the paper won the best research paper award.
In parallel, I learned about the upcoming Industrial Graduate School 
%in Reliable Embedded Sensor Systems (ITS ESS-H)
ITS ESS-H, and after several rounds of discussion, I started as an industrial doctoral student in 2017.

During this period,
Westermo migrated from a development model loosely based on scrum
to a feature-driven development model based on kanban, where
every feature is developed in isolation from the others
in separate code branches,
and then merged into a main branch after a risk management process.
%It is believed that this development model:
%(i) reduces risk, as the main code branches can be kept stable while
%giving developers freedom to do major changes in development branches,
%and also:
%(ii) it provides faster time to market since the main code branch
%now has the goal of always being ready for entering a release phase.
%\remarkP{Comment from Tom: What about integration testing at this point.}
%
This increased parallelization of the testing of the branches
makes it harder to get
resources for nightly testing, and makes the results more sparse,
in turn making the test results more complicated to understand.
As illustrated in Figure~\ref{20200312-growth-line},
the increase in complexity is not only visible in number of branches,
but also in the number of test cases and test systems in use, as well as
in the number of new test cases executed.

%\remarkP{Remove this paragraph?}
%I am also colorblind,
%a condition first described by Dalton in the late 1700s~\cite{dalton1798},
%and made famous in the Ishihara test~\cite{ishihara1918tests}
%where dotted plates in various colors are used to
%determine severity and type of color blindness.
%Being colorblind gives me a disadvantage with respect to visualizations,
%since a common approach in visualizations is to
%rely on presenting different types of information (e.g., pass, fail) with
%colors alone (e.g., red, green), which may be a pointless approach
%when viewed by a colorblind person, when printed or shown
%on through a projector with poorly calibrated colors~\cite{zeileis2009}.

My personal and industrial background %and position at Westermo
has given me unique insights
in industrial software testing, which may have had an impact on how
problems have been formulated, and how data has been collected and analyzed.
%\remarkW{also on problem formulation?}.

\mysection{Background}

\begin{figure*}[t]
\centering
\includegraphics[width=0.9\linewidth]{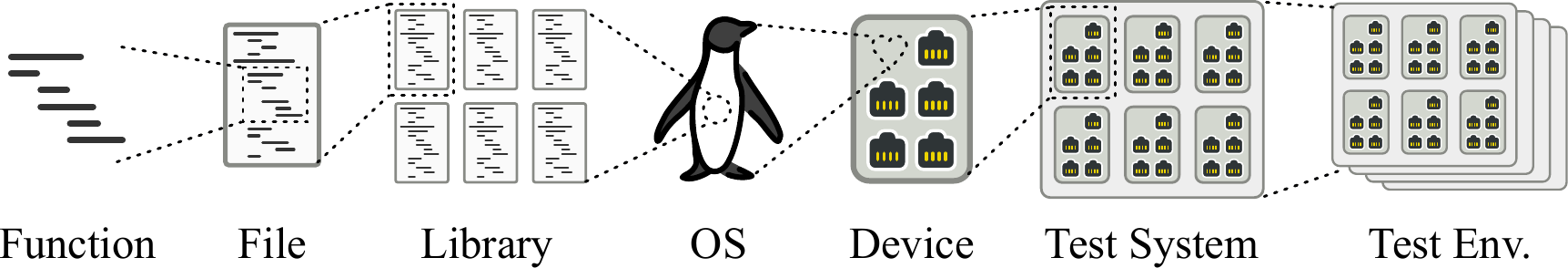}
\caption{
  Perspectives of testing:
  % Abstraction levels:
  from low level (source code functions)
  to high level (system).
  %\remarkD{Is this really a level of abstraction? Are there better words? Granularity? Integration?}
  %\remarkP{Perspective?}
\label{test-levels}
}
\end{figure*}

The growth in complexity in the nightly testing %at Westermo
(illustrated in Figure~\ref{20200312-growth-line})
is also present elsewhere, such as in the automotive industry. Staron \cite{staron2017automotive} reports on software architecture in this domain. There has been a growth in number of computers in a car, from at most one in the 1970s, followed by the introduction of electronic fuel injection, anti-lock brakes, cruise control in the late 1900s and autonomous driving and automated breaking when obstacles are detected in the early 2000s. Staron mentions that cars in 2015 could have 150 computers with more than 100 million lines of code, and that some of the trends here are: heterogeneity and distribution of software, variants and configuration, autonomous functions as well as connectivity and cooperation between nodes.

Embedded systems are becoming ubiquitous.
They range from portable sensors to
communication equipment providing infrastructure, as part of a train
or other vehicles, industrial plants or in other applications.
As many as 90\% of newly produced processors are part of embedded systems~\cite{garousi2017embedded}.
Software failures in communication equipment can lead to isolation of
nodes in a vehicle or a plant, leading to delays, loss of
productivity, or even loss of life in extreme cases.
The software in embedded systems
needs to be of high quality and software testing is the standard
method for detecting shortcomings in quality.

Software testing, can be
defined\footnote{This definition partially overlaps with
definitions from both the
International Software Testing Qualifications Board (ISTQB)
and the ISO/IEC/IEEE 29119-1 standard~\cite{iso291191, ISTQB-GLOSSARY}.
Many other definitions exist.
}
as the act of manually or automatically
inspecting or executing software
with or without custom hardware
in order to gather information for some purpose:
feedback, quality control, finding issues, building trust,
or other.
An important aspect of testing embedded systems is to do testing on real
hardware~\cite{banerjee2016testing,wolf1994codesign}.
For some of the testing, emulators and simulators can be suitable,
but these should be complemented with physical hardware~\cite{rosenkranz2015distributed}.
Testing could also iterate from being on host and target hardware~\cite{cordemans2014test}.
By testing on target hardware, timing and other non-functional aspects can be verified
along with functional correctness.
One common approach for testing on real hardware is to build test
systems of the embedded devices in network topologies,
in order to support testing.

%\remarkD{Very nice and mature up until this point. Not saying anything about the text below - I have yet to read it - but just thought I'd mention it.}

An overwhelming majority of the software testing
conducted in industry is manual. Kasurinen et al.\ found it as high as
90\% in a study in Finland in 2010~\cite{kasurinen2010software}.
A practitioner focused report from 2015
found that only 28\% of test cases
are automated~\cite{capgemini2015sogetti}.
%
% Furthermore, testing represents between 30\% to 80\% of the development cost.
%\remarkD{I know that this is a standard sentence to involve, but is it really meaningful?}
%\remarkP{removed.}
%
Test automation can reduce the cost involved and also improve time to market~\cite{wiklund2017impediments}.
If tests can run with minimal human intervention,
considerable gains in test efficiency are possible.
Indeed, many organizations strive for agile and continuous development
processes, where test automation is an important part,
and research has been done
on how to achieve this~\cite{olsson2012climbing}.
% step 1: agile process
% step 2: continuous integration with complete test suites
% step 3: to continuous deployment
For organizations developing networked embedded systems
to be successful in this transition, they would need a number of things
in addition to agile processes.
First of all, they would need a way to automate testing with
a test framework and test cases.
Test suites would have to run every now and then.
The test framework needs to know how to log in,
upgrade, and configure the hardware in the test systems.
If the organizations develop many different
hardware models, they would need
several test systems with different sets of hardware.
In order to utilize the resources optimally, the test systems
might be shared: humans use them by day
and machines run nightly testing when no one is in the office.
This way developers and testers could use the test systems
for development, fault finding and manual testing.
During the nights, the organizations would benefit from potentially
massive and broad automated testing.
This type of testing, on this high level, could be called
automated system level testing
in the context of networked embedded systems.
%This level of abstraction is
This perspective is
illustrated in the rightmost parts of Figure~\ref{test-levels}.

\begin{figure*}[t]
\centering
\includegraphics[width=0.7\linewidth]{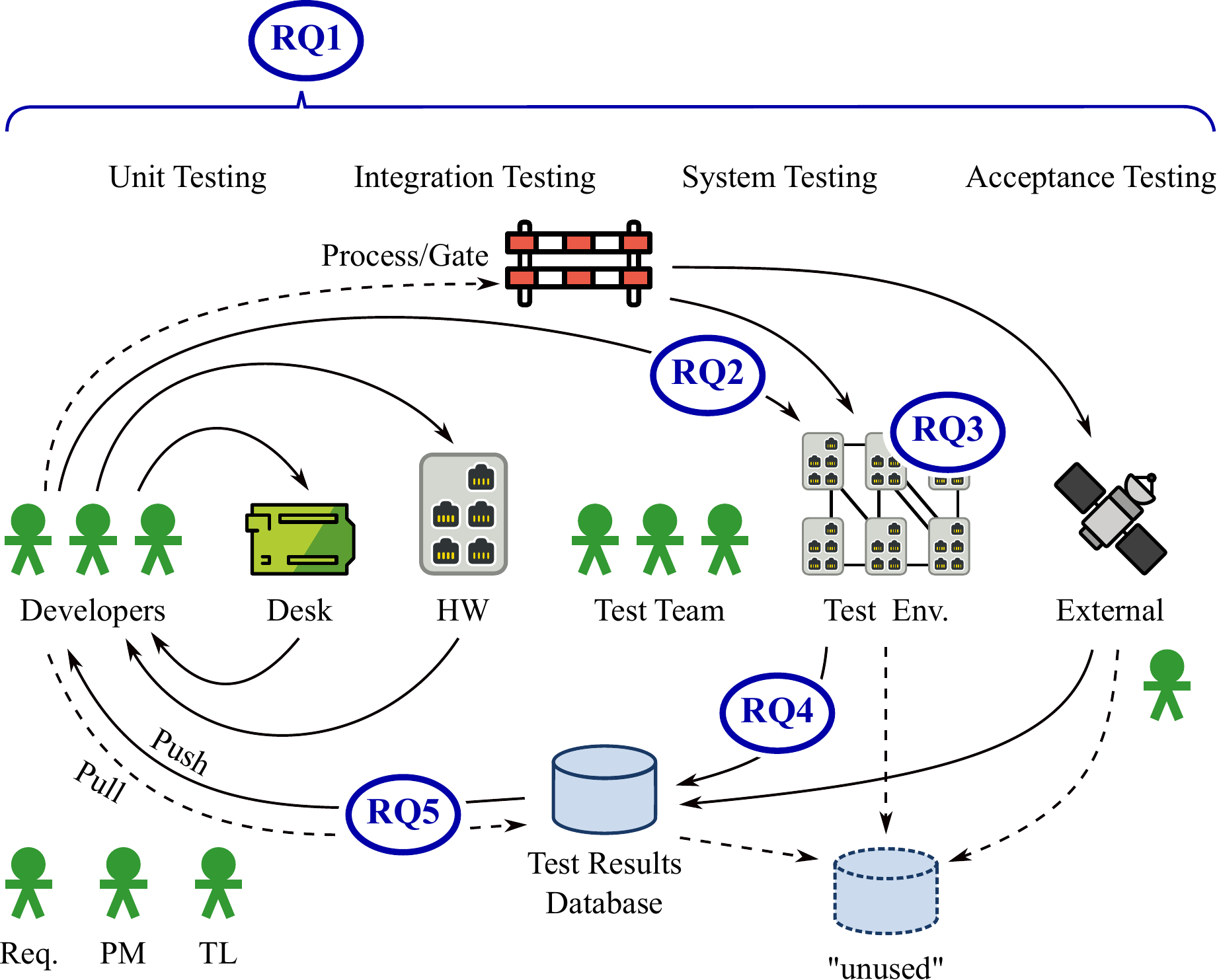}
\caption{
  Automated System-Level Software Testing of Industrial Networked Embedded Systems with highlighted research questions
  (compare Figure~\ref{fig-contrib-flow}). 
  \label{fig-trc-w-rq}
  }
\end{figure*}

Wiklund et al.\ identified lack of time for testing as an important
challenge for test automation~\cite{wiklund2017impediments}.
One reason could be that system level testing is slow
compared to unit testing where a function, file or library may be
tested (left part of Figure~\ref{test-levels}).
A typical test case for networked embedded systems
on system level could be a firewall test.
This test case would need at least three devices under test (DUTs): %\remarkD{Possibly move definition of DUT acronym here?}
a firewall node,
an internal node (to be protected by the firewall),
and an external node (trying to reach the internal node).
Depending on configuration and type of traffic, the outside node
should or should not be able to reach the inside node.
These test cases need time because they perform several configuration
and verification steps, send actual traffic in the network,
and the test framework analyzes the traffic sent.
%% Depending on how the test framework and the test cases have been
%% designed, each test case would have a number of configuration
%% and verification steps, in some order.
%% Tests that pass either trigger a pass criteria, or
%% do not trigger a fail criteria.
Because testing is slow, it may not be feasible to run all test cases
every night, so the organizations need to
decide on which test cases to include in or exclude from testing.
This problem of regression test selection
(RTS, or SL-RTS for System Level RTS)
is well-studied, but very little previous research had been done on SL-RTS
in 2016 when \paperissre{} was published.

Wiklund et al.\ also found that low availability of the test environment
is an important challenge for test automation~\cite{wiklund2017impediments},
in particular if the test systems have non-standard hardware~\cite{maartensson2016continuous}.
There is thus a need to maximize the value of the testing,
given available resources.
%\remarkP{Wasif suggests to add ``In Westermo's context,'' -- but I'm not sure I agree.}
%\remarkP{Think about how to generalize this.}
If a large number of physical test systems are built, and a large
number of test cases designed, then one would need a way
to assign the hardware resources of the test systems to the test cases,
so that as many test
cases as possible could run on as many test systems as possible.
In this thesis, this process is referred to as ``mapping'' of
test cases onto test systems.
If there is no automated mapping, then
each new test case might trigger a need to build a new test system,
which would be expensive and limit the scalability of the testing.
Also, in order to maximize the value of the testing, the mapping process
should make sure that, over time, each test case
utilizes different parts of the hardware in the test systems,
such that hardware coverage is increased.

A literature study from 2017 on continuous practices identified that,
as the frequency of integrations increase, there is an exponential growth
of information~\cite{shahin2017continuous}.
%The study also identified a related challenge:
%lack of awareness and transparency.
Similarly, Brandtner et al., identified that relevant information
might be spread out in several systems~\cite{brandtner2014supporting}.
One way to handle test results data is to implement a test results database (TRDB).
The TRDB could support exploration, visualization and also provide overviews of results, such that when
engineers come to work in the mornings, they could rapidly understand the results of the nightly testing.
% Log files, trend plots and other information should be readily available.
It is therefore important to study and learn from industrial
practitioners how the information flows in their
testing processes, and how they make decisions
based on visualizations and other support systems enabled by the TRDB.

Intuitively, testing might be expected to be a deterministic process --
given the same stimuli to the same system, one could expected the same response.
However, this is not true, or rather: 
in practice, no one is able to observe or control all variables
that are of relevance to a complex system, therefore testing is perceived as non-deterministic when some tests intermittently fail.
This has been well studied, e.g.\ by Luo et al.~\cite{luo2014empirical}
and has been shown to be linked to
poor quality test code~\cite{garousi2018smells}.
However, most of the previous work has been conducted on unit level testing of open source software.
It appears as if only four previous studies have considered higher level testing~\cite{ahmad2021empirical, eck2019understanding, lam2019root, thorve2018empirical}.

%%%%%%%%%%%%%%%%%%%%%%%%%%%%%%%%%%%%%%%%%%%%%%%%%%%%%%%%%%%%%%%%%%%%%%%%%
%\newpage
\mychapter{Research Process}
\label{kappa-research-process}

% --- ---  --- --- ---  --- --- ---  --- --- ---  --- --- ---  --- --- ---
% New 2020-08-20
The research of this thesis has largely followed the process described in Figure~\ref{fig-collab}.
The common problem has been that of automated system-level software testing of industrial networked embedded systems. The work has been funded by Westermo and the Swedish Knowledge Foundation through the ITS ESS-H research school,
and collaborative work has been done, primarily between the Software Testing Laboratory at Mälardalen University and Westermo. Through the research effort new knowledge has been generated, which gives benefits and value to the company, the researchers, and for society in general. The knowledge has led to improvements in work processes and tools used at Westermo, and to academic publications. The value and benefits have led to an interest in continued collaborative research.

% --- ---  --- --- ---  --- --- ---  --- --- ---  --- --- ---  --- --- ---
% New 2020-09-09
The \textbf{research goal} of this thesis is:
\emph{to improve automated system-level software testing of industrial networked embedded systems.}
The five research questions of this thesis
are organized to align with Figure~\ref{fig-trc-w-rq}, which shows
the overall information flow model discovered in \papertrc.
In short, this thesis targets five questions that can be positioned in different stages of this flow model:
How does information relevant to testing flow in an organization developing embedded systems?
Which test cases should be executed when there is not enough time?
Which hardware devices should be included in the testing when there are too many possible combinations?
Why do some test cases intermittently fail?
How can information relevant to testing be presented to developers and testers such that they can make decisions?
%
% --- ---  --- --- ---  --- --- ---  --- --- ---  --- --- ---  --- --- ---
The rest of this chapter describes these five questions in more detail,
the case study research method,
as well as a discussion on research quality.

% \section{Research Goal and Research Questions}
\mysection{Research Questions}
\label{my-research-goals}
\label{my-rqs}

A challenge with testing embedded systems is understanding
how updated software reaches a test environment,
how testing is actually conducted in the test environment,
and how test results reach the developers that updated the software in the first place
(i.e.\ the elements and arrows in Figure~\ref{fig-trc-w-rq}).
The following research question (RQ) was formulated in order to tackle this problem:

\begin{description}
\item[RQ1:]{%
  How could one describe the flow of information in software testing in an organization
  developing embedded systems, and what
  key aspects,
  challenges,
  and
  good approaches
  are relevant to this flow?
  (\papertrc)
}
\end{description}

\noindent
Despite a long history of research, the regression test selection problem was relatively
unexplored from a system-level perspective at the time of \paperissre{} (2016).
In Figure~\ref{fig-trc-w-rq}, test selection is illustrated as a process occurring before testing and before software reaches the test environment.
The following RQ was formulated in order to explore this problem from a system-level:

\begin{description}
\item[RQ2:]{%
    What challenges might an organization have with respect
    to system-level regression test selection in the context of
    networked embedded systems,
    and how could one address these challenges?
    (\paperissre)
}
\end{description}

\noindent
Similar to how regression test case selection occurs prior to the actual testing, a subset of the available hardware must also be selected for each test case -- the \emph{mapping} of requirements of the test cases onto test systems.
In Figure~\ref{fig-trc-w-rq}, a mapping can be thought of a process requiring a test environment.
This leads to the following RQ:

\begin{description}
\item[RQ3:]{%
    What challenges might an organization have with respect
    to test environment assignment in the context of
    networked embedded systems,
    and how could one address these challenges?
    (\papermapper)
}
\end{description}

\noindent
When system-level test cases are executed over time, night after night, on embedded systems in an industrial setting,
there is almost always a drift in the software, testware or hardware between test sessions.
With the exception of mostly manual debugging and trouble shooting, rarely do companies invest effort in doing regression testing of unchanged systems.
Instead, one desires confirmation that recent changes have not introduced regressions.
Under these conditions, test cases may sometimes intermittently fail for no obvious reason,
leading to unexpected test results over time in the test results database (Figure~\ref{fig-trc-w-rq}).
This problem motivates the following RQ:

\begin{description}
\item[RQ4:]{%
    What are the root causes of intermittently failing tests during system-level testing
    in a context %where the environment is
    under evolution,
    and could one automate the detection of these test cases?
    (\paperflaky)
}
\end{description}

\noindent
System-level software testing of industrial networked embedded systems produces large amounts of information, and involves many automated decisions.
In the software development process, the results may be hard to render into something actionable,
in particular in the daily activities of a software developer,
when a new feature is about to be completed,
and at release time.
%If a developer expects a test to have failed and queries the test results database (Fig.~\ref{fig-trc-w-rq}) without finding it,
%it can be hard to understand if the test is not shown in the list of failures because it passed,
%because it was not included in nightly testing, or because it could not be mapped to the test systems being used.
%
%
%\remarkP{Focus on the three steps in the development process instead?}
%\remarkP{Comment from Daniel: think about type-errors -- does the answer match the question?}
%When large-scale system-level software testing of industrial networked embedded systems has been conducted,
%massive amounts of information has been created, and many automated decisions have been made.
%When a developer comes to work in the morning, he or she must make this information actionable,
%e.g.\ by understanding if a verdict from a certain test case is not shown because it passed or because it was not included in nightly testing.
This leads to the final RQ of this thesis:
% \remarkD{Ok, still not perfectly happy with this last RQ, but it's ok.}
% \remarkP{Yes.}

\begin{description}
\item[RQ5:]{%
  How could one implement and evaluate a system to enhance
  %decision-making,
  visualization and exploration of test results to support the information
  flow in an organization? % \remarkW{last part involving information flow necessary?}?
  % \remarkP{Use cases?}
  (\papertim)
}
\end{description}

% \mynewpage
\mysection{The Case Study Research Method}

%
% \subsubsection{Case Studies}
%

%\remarkP{Wasif suggests to italicize quotes, but since we don't in the Tim paper, or in the info-flow paper, I don't think we should -- discuss?}
According to Runeson et al.~\cite{runeson2012}, case study research in software engineering is useful in order to understand and improve software engineering processes and products.
They define the case study research method as: 
\begin{quote}
an empirical enquiry that draws on multiple sources of evidence to investigate [\ldots]~a contemporary software engineering phenomenon within its real-life context, especially when the boundary between phenomenon and context cannot be clearly specified.
\end{quote}

The studies in this thesis are all case studies,
but differ in terms of aim
%focus\remarkD{, aim(?)}
and type of data used -- both qualitative and quantitative data has been collected and analyzed.
%\remarkW{See one of previous comments in 2.2 where you say a large portion is qualitative.}
%\remarkP{I cleaned up a little bit, should be OK now.}
\papertrc{} exclusively used qualitative data (in the form of interviews),
Papers~\pmapper{} and \pflaky{} primarily used quantitative data,
and Papers~\pissre{} and \ptim{} combine qualitative and quantitative data.

%
% \subsubsection{Qualitative Data Collection and Analysis}
%
When \papertrc{} was being planned, 
%When planned the interviews for \papertrc{},
a number of guidelines were considered.
Kitchenham and Pfleeger published a series of guideline papers in 2001 and
2002~\cite{pk-part-3,pk-part-4,pk-part-5,pk-part-6,pk-part-1,pk-part-2}.
These had an influence on, for example,
Runeson et al.~\cite{runeson2012},
that in turn has had an influence on the overall guideline
paper followed for \papertrc: Linåker et al.~\cite{linaker2015}.

A total of about 28 hours of audio and video has been recorded 
and transcribed into about 185 pages of text for this thesis
(17 hours of audio transcribed into 130 pages of text for \papertrc,
7 hours of video transcribed into 36 pages of not as dense text for \papertim, 
and 4 hours of interview audio and 20 pages of text was collected for \papertrdb, on which \papertim\ is based).
%, we recorded another 4 hours and got 20 more pages of textual data.
For the data analysis of this qualitative data, many options were available, e.g.\ 
grounded theory (for example described by
Stol et al.~\cite{stol2016grounded}),
content analysis (as described for the field of nursing by Graneheim
and Lundman~\cite{graneheim-lundman-2004}),
and thematic analysis.
There is a significant overlap between these methods,
and also a drift of guidelines into software engineering from the social sciences
(e.g., the Cruzes and Dybå guidelines for thematic analysis~\cite{cruzes2011recommended}).
In the initial stages of data analysis for \papertrc,
we held a workshop among the authors where
the decision to use the recommendations of thematic analysis
described for the field of psychology by Braun and Clarke~\cite{braun-clarke-2006} was taken.
Among the authors we saw this approach as easy to understand and suitable for the type of data at hand.

At the time, I (and the other authors too) felt that the integrity and anonymity of the interviewees was of great importance.
The interviewees felt the same, and some of them pointed out
that leaked transcripts could uniquely identify them, or be a form industrial espionage.
As part of the data collection process of \papertrc{},
an anonymization process based on existing ethical guidelines, e.g.~\cite{CODEX2}, was designed.
This, in turn, lead to \paperethics{} on ethical interviews in software engineering.

%
% \subsubsection{Quantitative Data Collection and Analysis}
%
At the core of \paperissre{} is an algorithm and tool for SL-RTS
that was evaluated with data from four years of nightly testing --
two years of data from before introduction of the tool and two after.
\papermapper{} proposes and describes the implementation of an algorithm and tool for hardware selection, which was evaluated
using 17 test systems and more than 600 test cases
for a total of more than ten thousand possible mappings.
This way, performance of the new tool could be compared with the previous tool in use at Westermo.
\paperflaky{} describes a novel algorithm to identify intermittently or consistently failing tests.
The algorithm was used on test results from nine months of nightly testing and
more than half a million test verdicts were used in order to identify test cases
to investigate further.
The extended analysis used project artifacts, such as code commit messages, bug trackers and so on, in order to find patterns for why test cases are intermittently failing.
Finally, \papertim{} investigates how to implement visualizations based on test results, which again revolves around the development of a tool.
The tool was evaluated by transcribing recorded reference group meetings, by member checking and by analyzing log files from using the tool.

%\remarkD{Best method section in a collection of paper thesis I have ever read. It is detailed and meaningful.}
%\remarkP{:-)}
%\remarkW{Yes, I can second, mainly that it is not boring to read :-)}

\mysection{On Research Quality}
% \section{Replicability, Rigor, Relevance, Generalizability and Research Ethics}

%\remarkP{2021-07-07: This section is a bit of a mess -- I like the text, but it needs cleanup.}

The term \emph{software engineering} was coined by Margaret Hamilton while working on the Apollo program at NASA \cite{hind2019first} in an attempt to bring discipline, legitimacy and respect to software.
As pointed out by Felderer and Travassos, software engineering has evolved
from its seminal moments at NASA (and NATO) aiming at observing and understanding software projects in 1960s and 1970s,
via attempts to define methods for empirical studies in the 1980s
(such as the work on goal-question-metric by Basili and Weiss \cite{basili1984methodology}),
%
%\remarkD{Maybe consider the work of Victor Basili here as well?},
to the formation of publication venues for empirical software engineering in the 1990s,
into an evidence-based field of research~\cite{felderer2019evolution}.
% \remarkW{unless you put references, this text is too much to trust.}
% \remarkP{I moved the Felderer and Travassos reference to end of sentence.}
In 2012, Runeson et al.\ published a book called ``Case Study Research in Software Engineering -- Guidelines and Examples''~\cite{runeson2012}
which is one of the main guidelines used in this thesis.

As mentioned above, Arora et al.\ \cite{arora2020changing}, argues that one of the strengths of 
research at companies is the access to data.
However, when a company uses its own data and tools to make research claims such as in this thesis, e.g.\ by using the SuiteBuilder tool ``two thirds of the failing tests are now positioned in the first third of the test suites,'' then this claim cannot easily be confirmed or rejected by someone not at the company. The term falsifiability was coined by Popper in 1934 \cite{popper2005logic}, and it is sometimes used as a possible demarcation between science and pseudoscience.
Without access to company data and tools, would it be any easier to falsify the statement about distribution of failing tests, than it would be to falsify the claim that a teapot orbits the Sun somewhere between Earth and Mars (Russell's teapot)? Intuitively, one might want to argue that this thesis is scientific, but that belief in Russel's teapot is pseudo-scientific. 
Furthermore, in his famous paper ``Why Most Published Research Findings Are False'' \cite{ioannidis2005most}, Ioannidis argues that lack of replication, lack of publications with negative results, and over-emphasis on statistical significance are root causes for why many research findings are false.
Again, the statement on the fail distributions is not only hard to falsify, it is also hard to replicate. 
Claims that are easy to replicate are of course easier to falsify.
%
%\remarkP{Discuss this sentence?}
In retrospect, now in 2021, one would argue that we as authors of the papers in this thesis, should have made a greater effort to make data and tools available to a greater audience. 
%In retrospect, now in 2021, one would argue that I (or we as co-authors, or we as a research community), for all the paper in this thesis, should have made a greater effort to make data and tools available to a greater audience. 
Or that we, in addition to exploring challenges at one or a few companies, would also have used publicly available sources such as open source projects to evaluate findings, tools or hypotheses.  
If we had, then the opportunities to replicate research findings and the generalizability of the findings would have been better.
%
%\remarkP{2021-08-10: Is the above good enough for this comment by KL: "Does this mean other researchers cannot repeat? Add a few lines on this in the kappa."}

% \remarkP{CLEAN THIS UP A BIT -- NEEDS MERGE WITH OTHER SUBSECTIONS?}

%\remarkP{Margaret Hamilton coined the term while at NASA.}
It has been suggested that good research should have both rigor and relevance. By doing so, one avoids conducting research only for the sake of researchers.
%In the coming paragraphs, we also discuss ethical aspects of research, as well as context and generalizability.
%
At the core of \emph{rigor} are carefully considered and transparent research methods~\cite{ivarsson2011method,runeson2012,sanno2019increasing,vermeulen2005rigor}.
%
% In order to have \emph{rigor} in the research of this thesis, many guidelines for the research methods, have studied and followed.

%Providing information on \emph{context} in software engineering research is relevant in order to allow readers to draw valid conclusions from
%the research~\cite{dybaa2012works,pw-context-2009,runeson2012}.
Doing research in a certain context certainly has an impact on \emph{generalizability}.
However, Briand et al.~\cite{briand2017case} argue that
``we see the need to foster [\ldots]\ research focused on problems defined in collaboration with industrial partners and driven by concrete needs in specific domains [\ldots]''
In other words, research in a narrow context may have high \emph{relevance}, regardless of its generalizability.
%\remarkW{Perhaps this paragraph will be more suitable to go as the ending paragraph for 2.2.}
%\remarkP{Yes, good suggestion -- I moved half of it there.}
%
% In the next subsection we describe the context for this.

Research involving human practitioners may harm these individuals, and
\emph{ethical aspects} are of relevance to businesses as well.
Smith \cite{smith2017ethical}\ points out that being an industrial doctoral student comes with
unique ethical risks, i.e.\ an industrial doctoral student might portray his of her employer in a favorable light.
%
%\remarkP{Rewritten without I:}
For the research in this thesis, we have had a focus on minimizing harm, which has led to guidelines not part of this thesis (in papers \pethics\ and \pairts~\cite{strandberg2019ethical, strandberg2021ethical}).
In addition to avoiding harm to participants in the research, another ethical aspect of this thesis is the potential
for hiding decision making criteria of an algorithm, which risks making decisions inscrutable, and thereby disempowering stakeholders~\cite{martin2019ethical}.
%\remarkP{\textbf{Remove?}
%During the work in this thesis, I have aimed at minimizing harm.
%I have also proposed guidelines for how others could consider ethical aspects when involving industrial practitioners,
%both as a single author and with others (in papers \pethics\ and \pairts~\cite{strandberg2019ethical, strandberg2021ethical}).
%In addition to causing harm to participants in the research, another ethical aspect the potential for hiding decision making criteria of an algorithm, thereby making decisions inscrutable, which may disempower stakeholders~\cite{martin2019ethical}.
%}

%\subsection{Validity Analysis}
%
% HESS 6
% Acknowledge the study’s limitations
%
% HESS 4
% Consider alternative explanations of the findings
%
% TODO: ???

% Conclusion validity
%  - Does the treatment/change we introduced have a statistically significant effect on the outcome we measure?
% Internal validity
%  - Did the treatment/change we introduced cause the effect on the outcome?
%  - Can other factors also have had an effect?

%\remarkP{Is my discussion of mixing qual and quan BS?}
% internal validity
One challenge with the case study research method in software engineering is that when exploring
a ``phenomenon [\ldots]\ when the boundary between phenomenon and context cannot be clearly specified''
(as Runeson et al.~\cite{runeson2012} defines a case study),
the findings might depend on % come from \remarkD{change to "depend on"?} 
the context and not the phenomenon.
In industrial case studies, many variables may have an impact on what is being measured.
There could therefore exist confounding factors, perhaps changes in a software development process,
that have not been investigated, or other threats to \emph{internal validity}.
However, this threat has been addressed by involving both qualitative and quantitative research methods,
and by collecting data as to minimize the impact of confounding variables.
E.g., in \paperissre, challenges with respect to regression test selection were identified qualitatively.
\paperflaky{} used a quantitative metric for test case intermittence that led to the
identification of test cases that were then qualitatively investigated.
Furthermore, prolonged involvement and industrial experience should also have addressed this threat.
% conclusion validity -- skip!

% Construct validity
%  - Does the treatment correspond to the actual cause we are interested in?
%  - Does the outcome correspond to the effect we are interested in?
In the studies in this thesis, a number of constructs are used
(e.g., a flaky test, the mapping problem, and information flow).
These do not always have a standard definition, which leads to threats to
\emph{construct validity}, not only in the research in this thesis, but in general
in the research community.
One way to combat this threat to validity is to carefully define terms, and be transparent about
algorithms used to collect data and also how qualitative data has been analyzed.

% External validity, Transferability
%  - Is the cause and effect relationship we have shown valid in other situations?
%  - Can we generalize our results?
%  - Do the results apply in other contexts?
In all but one of the papers of this thesis, the results stem from one context:
the automated system-level testing at Westermo.
One should therefore consider \emph{external validity} before transferring results into other contexts.
However, there is a growing body of knowledge of industrial testing and system-level testing.
It is also interesting to speculate about generalizability, e.g.\ on test selection or
intermittently failing tests, from unit-level testing to system-level testing.
One such example could be the finding in \paperflaky{} that test case assumptions (e.g., on timing) is a factor for intermittence that seems to be common for both unit- and system-level testing.

% Credibility
%  - Are we confident that the findings are true? Why?
% Dependability
%  - Are the findings consistent?
%  - Can they be repeated?
% Confirmalibity
%  - Are the findings shaped by the respondents and not by the researcher?
By being transparent in the choice of, and strict in the use of, scientific methods, % of the research in this thesis,
while at the same time considering humans and organizations,
the research in this thesis should be not only rigorous and relevant, but also ethical.

%\remarkD{Daniel is here in review}

% \remarkP{Harmonize with I-A Collaboration section.}

%%%%%%%%%%%%%%%%%%%%%%%%%%%%%%%%%%%%%%%%%%%%%%%%%%%%%%%%%%%%%%%%%%%%%%%%%

%\newpage
\mychapter{Related Work}

\label{previous-research}

%\subsection{Summary}

The research fields of software testing,
regression test selection,
intermittently failing tests
as well as communication and visualization
are all well studied.
However, there is a notable gap between academia and industry in these fields:
an overwhelming majority of the actual testing in the industry is manual,
%test cases are designed based on experience instead of using state of the art methods,
%regression test cases are selected with poor methods,
%and most research is evaluated with small scale examples and not in a industrial settings.
%and there is also suspicion between
%the two worlds of agile testing and testing for safety critical software.
%
%
despite more than 40 years of research on regression test selection
very little of the research targets the system level,
%
%\remarkP{Remove ISTQB from related work?}
%Furthermore,
%test organizations like the ISTQB have recently taken an interest
%in software test automation and have produced very relevant guidelines
%for the field of software testing.
%
intermittently failing tests have almost exclusively been explored on unit level testing of open source software,
and previous work on the role of communication in software engineering and
software testing indicate that a requirements specification is of
great importance -- in practice these are unfortunately typically of poor quality.
%% \footnote{%
%% %
%% The lack of evaluated methods on industrial systems does not only seem to be present in the area of software testing. In a systematic mapping study in the field of reliability, Pietrantuono investigated literature on the test resource allocation problem~\cite{pietrantuono2019testing}.
%% He found that most approaches for allocating test resources were never evaluated in practice, instead they were evaluated with ``numerical illustrations and artificial datasets.'' %\remarkW{Move to footnote?}
%% %\remarkP{Yes.}
%% }
%
%
%
%When doing research in software engineering and visualizations
%one must consider human factors,
%cognitive biases, as well as disabilities like color-blindness.
%
%
This chapter discusses findings in these areas of research, and also briefly
mention the mathematical field of
graph theory and the subgraph isomorphism problem -- topics
related to \papermapper.

\mysection{Industrial Software Testing}

A lot of research has shown that industry seems to be slow to adopt
state of the art techniques for software testing.
Most testing is done manually, perhaps as much as 90\%~\cite{kasurinen2010software}.
%Even though manual testing is very common there are no secondary
%studies on how manual testing is conducted~\cite{garousi2016}.
Well-defined test design techniques exist, but testers
``rely on their own experience more''~\cite{illes2008role}.
%
%``[A]pproaches to test case selection are usually
%oriented towards [either a] risk-based selection
%[or a] design-based selection\ldots{}''~\cite{kasurinen2010test}.
%
%The oracle problem is also a challenge to testers
%(a test oracle is ``a procedure that distinguishes
%between the correct and incorrect'')
%\cite{barr2015oracle}.
%
%Traceability is typically used between
%requirements and test cases.
%Traceability between code and test cases
%can be desirable but can be hard to get
%for system-level testing~\cite{spieker2017reinforcement}.
%Traceability between issues and code, and how to
%automatically retrieve them when lost, has been studied~\cite{zhang2015survey}.
%
The industrial use of issue trackers\footnote{Issue tracking can also be referred to as
bug tracking, %~\cite{zhang2016literature},
defect reporting, %~\cite{ISTQB-FL},
or incident reporting~\cite{iso29119, ISTQB-FL, zhang2016literature}.
%~\cite{iso29119}
%\remarkD{issue trackers should be a commonly known term, but I am not sure it is...}
%\remarkP{done}
}
is sometimes flawed, and many organizations don't use one~\cite{kasurinen2010software}. %issue trackers
%\remarkP{KL: People know about the importance of requirements, but industry doesn’t do this. Why?}

On the positive side, companies change over time and many strive towards
using more agile practices, such as shortening feedback loops and
adopting continuous integration~\cite{olsson2012climbing}.
%%
%%
%% There seems to be two camps in the world of software testing:
%% one camp that wants to move towards agile and iterative practices,
%% for example by performing exploratory testing.
%% %% In~\cite{kasurinen2010test} the authors found that
%% %% ``the organizations were strongly
%% %% divided into two opposing groups; some organizations considered
%% %% explorative testing as an important phase where usability and user
%% %% interface issues were addressed, whereas some organizations considered
%% %% testing without test cases and documentation as a waste of test
%% %% resources.''
%% Afzal et al. showed that exploratory testing can be as good as, or better,
%% than conventional test case based testing~\cite{afzal2015experiment}.
In a literature study on continuous practices from 2017, Shahin et al.\
identified that, as code integrations become more frequent,
the amount of data such as
test results will increase exponentially~\cite{shahin2017continuous}.
Therefore it is ``critical to collect and represent the information
in [a] timely manner to help stakeholders to gain better and easier
understanding and interpretation of the results\ldots''
%Also: ``[CI tools] produce huge amount of data
%that may not be easily utilized by stakeholders\ldots''
%Continuous integration is an important agile practice.
There is also a need for speeding up continuous integration~\cite{spieker2017reinforcement}.
%highlights the need
%and~\cite{shahin2017continuous}
%identified challenges for continuous integration.

%% The second and more traditional camp is common in publications
%% on safety-critical systems.
%% This research has typically focused on
%% structural or requirements-based coverage analysis for
%% functional, model-based, or structural testing~\cite{barbosa2011software}.
%% \cite{ghanbari2016seeking}
%% shows that many companies doing safety development
%% do not follow the strict processes.

%\remarkP{WA: Does this paragraph fit in?}
Agile software development practices can coexist with traditional approaches,
and they seem to benefit from each other.
%% However, these two camps may co-exist,
Notander et al.\ points out that a
``common belief is that agile processes are in conflict with the
requirements of safety standards\ldots\ Our conclusion is that this
might be the case [sometimes], but not [always]''~\cite{notander2013challenges}.
Ghanbari came to similar conclusions~\cite{ghanbari2016seeking}.
A study on communication in agile projects
suggest that plan-driven approaches are sometimes needed
in agile contexts~\cite{pikkarainen2008impact}.
%``While the use of [agile practices] facilitate
%[some communication] the use of agile practices requires that
%the team and organization use also additional plan-driven practices to
%ensure the efficiency of external communication\ldots''
Similarly, a recent study by Heeager and Nielsen
identified four areas of concern
when adopting agile practices in a safety critical context
(documentation, requirements, life cycle and testing)~\cite{heeager2020meshing}.
% \remarkW{Nice but the connection of this para with industrial sw testing is bit unknown.}
% \remarkP{Yes, slightly rephrased now.}
%\remarkP{KL: People know about the importance of requirements, but industry doesn’t do this. Why?}
%\remarkP{2021-07-07: Discuss with Daniel and Wasif -- there seems to be some research on agile and RE, but not much empirical work?}

When it comes to testing of embedded systems,
an important aspect is to investigate
non-functional qualities such as timing,
and by testing on real hardware one can achieve this~\cite{banerjee2016testing,garousi2017embedded}.
One could also alternate between testing on hardware,
virtual hardware and no hardware~\cite{cordemans2014test}.

In a literature study from 2019, bin Ali et al.\ involved practitioners in
the research process in order to identify
industry-relevant research on regression testing~\cite{bin2019search}.
Among other things, they recommend researchers to:
evaluate coverage of a technique at the feature level (other coverage metrics were not seen as relevant for practitioners),
report on relevant context factors in detail as opposed to reporting generally on many factors
and to study people-related factors.

\mysubsection{Regression Test Selection}

%\remarkD{Narrative or not here? Discuss... Actually goes for entire section 3}

\label{previous-research-rts}
When organizations move towards nightly testing,
and run testing on the embedded systems,
they risk ending up with nights that are not long enough --
too many test cases and too little time.
This is a strong motivator for introducing
regression test selection (RTS).
A well-cited paper by Yoo and Harman proposes three
strategies for coping with RTS: %regression test selection:
\emph{Minimization},
\emph{Selection},
and \emph{Prioritization}~\cite{yoo-harman2012}.
RTS can be based on many different properties:
code coverage~\cite{mondal2015exploring},
expected fault locations~\cite{ostrand-weyuker2005predicting},
topic coverage~\cite{hemmati2015prioritizing}, or
historic data such as last execution, fault detection, and coverage data~\cite{elbaum2014techniques,engstrom2011improving, hemmati2015prioritizing,kim2002history}.
Mathematical optimization approaches
%targeting for example
%optimization on cost, code coverage, test diversity and/or test duration
have been used~\cite{fischer1977, herzig2015art, mondal2015exploring} as well as genetic algorithms~\cite{walcott2006timeaware}.
% \remarkP{Find citation on machine learning and RTS?}
Software fault prediction is a related field, where one
influential paper was written by Ostrand et al.\ for traditional
software development in large software systems~\cite{ostrand-weyuker2005predicting}.
%, it focuses on
%identifying where in the source code base faults are likely to appear. An
%influential paper by Ostrand et al.\
%investigates large software systems, but focuses on manual testing and
%making recommendations once per release,
%as opposed to the more frequent recommendations one could expect in
%continuous integration environments
Elbaum et al.\
mentions that traditional regression testing
techniques that rely on source code instrumentation and availability
of a complete test set become too expensive in continuous integration
development environments, partly because of the high frequency of code
changes~\cite{elbaum2014techniques}.
% makes code coverage metrics imprecise and obsolete.
%% \cite{saff2003reducing} consider several strategies for
%% continuous testing targeting a
%% reduction of non-productive time spent by developers.
%% \cite{marijan2015multi} described a study that used impact of
%% detected failures, test execution time and failure frequency of a test
%% case and functional coverage to prioritize test cases
%% efficiently in the context of continuous integration.

In 2017, when preparing my licentiate thesis, I identified four recent surveys on RTS, none of which was applicable to system-level RTS.
%\begin{enumerate}
%\item
The most recent one,
  from 2016 by Hao et al., mentions
  that most techniques have been evaluated with
  programs smaller than 6 kSLOC (SLOC = source lines of code)~\cite{hao2016test}.
  %and considers projects of about 10 kSLOC a reduction to this shortcoming
  This is still very far from many embedded systems where
  using the Linux kernel alone results in a code base
  in the order of 10 MSLOC.
%
%\item
Second,
  Catal and Mishra found
  that the dominating techniques for RTS are coverage based~\cite{catal2013test}
  % --
  % however %, these approaches are not applicable on a system level
  %since the instrumentation required will interfere
  %with the system under test.
  and these can therefore be difficult to apply to system-level software testing
  due to the instrumentation required.
%
%\item
The third survey, by
  Yoo and Harman, % published a well cited study in 2012.
  %They
  discuss a design-based
  approach to run what seems to be unit-level test cases:
  ``Assuming that there is traceability between the design and regression
  test cases, it is possible to perform regression test selection of
  code-level test cases from the impact analysis of UML design models''~\cite{yoo-harman2012}.
  %
  %We speculate that this type of traceability
  %is not so common in industrial agile environments.
  %\remarkP{Why do we speculate about this?}
%
%\item
  Finally, Engström et al.\ published a paper in 2010
  in which they investigated 28 techniques, where
  14 were on statement level, the others were no higher than module level~\cite{engstrom2010systematic}.
%\end{enumerate}

Now, in 2021, the body of knowledge on system-level RTS has grown and is still growing.
%A few of the more recent papers (some of which cite \paperissre) are:
%(i)
Ricken and Dyck did a survey on multi-objective regression test optimization in 2017~\cite{ricken2017survey}
and they argue that duration needed for test cases should be taken into account.
%as a prioritizer in the SuiteBuilder tool. \remarkD{Previous sentence. Is SuiteBuilder a recognizable term here?}
% \remarkP{They were actually discussing our ISSRE paper. Slightly rephrased now.}
%(ii)
Lou et al.~\cite{lou2019survey}
did a literature study on regression test prioritization.
They found that most approaches are not much better than simple ones,
the approaches target domains where testing is so fast that selection is not really relevant, and that no free and suitable tool exists (e.g.\ a tool integrated with jUnit).
%(iii)
In a 2020 doctoral thesis on test selection, Haghighatkhah~\cite{haghighatkhah2020thesis}
proposes a RTS approach that combines diversity (using test case distances) and
historic test data.
Similarly, diversity has also been proposed by Neto et al.~\cite{neto2018visualizing}.
%(iv)
In a systematic mapping study from 2020 on test case prioritization in continuous integration environments, Prado Lima and Vergelio
found that history-based approaches is a dominating approach in
recent publications on RTS, few of the publications they processed report on challenges in testing
in a continuous integration context and that
future work should also include aspects of intermittently failing tests (flaky tests) as well as time constraints~\cite{lima2020test}.

\paperissre{} builds upon the existing body of knowledge in the field of RTS, as it was in 2016, and the paper
shows feasibility of system-level RTS by using a framework of prioritizers that each
investigate some priority-giving aspect.
%This has contributed to the body of knowledge on system level RTS of 2020 \remarkW{this reads bit awkward to me, may be you want to say that even in 2020 more studies like SuiteBuilder needs to be done?}. \remarkW{Overall, the coverage of review papers in this section is superb!}
%\remarkP{Our paper on system level RTS, most are still not on SL. Bridge 2016 and 2020.}
%
% \remarkP{Does this fit here?}
% In our research we have extended the academic body of knowledge
% on this topic with \paperissre{} and findings related to system level RTS
% for networked embedded systems.
%
% \remarkP{What would we do differently if we re-implemented SuiteBuilder?}
Had the SL-RTS been implemented from scratch today, then one could have also considered the
diversity of the selection (see, e.g., Haghighatkhah~\cite{haghighatkhah2020thesis}),
and made the tool more scalable as to avoid making a test results database a bottleneck
when test complexity increases (see Figure \ref{20200312-growth-line}).
The SuiteBuilder tool described in \paperissre, has had its database questions simplified a few times in the seven years that have passed since its implementation. A migration from the old database schema and programming language has started, as described in \papertim{}.

\mysubsection{Test Environment Assignment}

The body of knowledge with respect to assigning
test environments to test cases for execution
seems to be in its infancy.
While \papermapper{} was in submission, Kaindl et al.\ published work on the same high level problem.
In their paper they ``briefly sketch the technical essence of [their] project'' explaining that they plan to use a semantic specification,
an ontology, and a taxonomy to assign test environments to a test case~\cite{kaindl2018verification}.
However, it seems as if \papermapper{} is the only study with a working solution for this problem in the domain of networked embedded systems.

\begin{figure*}[t]
\centering
\includegraphics[width=0.8\linewidth]{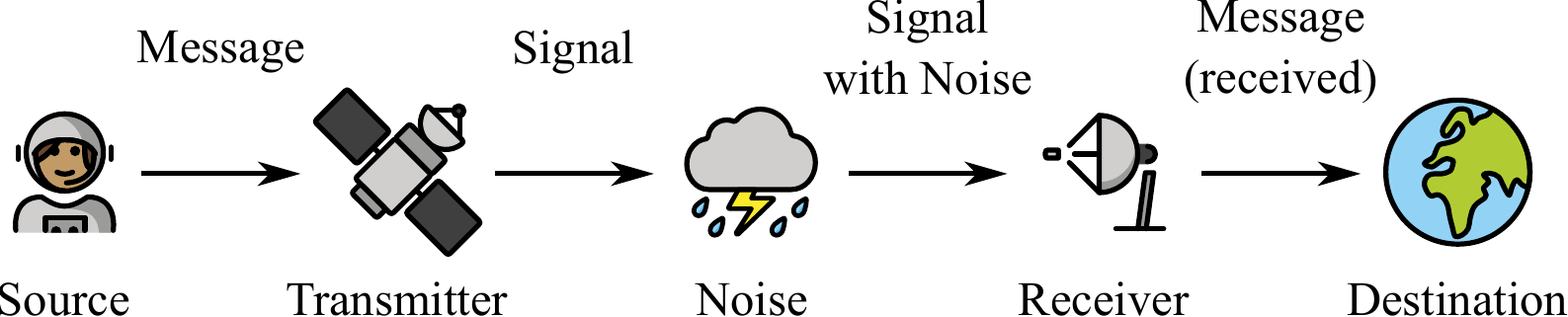}
\caption{
The Shannon-Weaver communication model.
%(Image
%in public domain, from
%\href{https://commons.wikimedia.org/wiki/File:Shannon_communication_system.svg}{Wikimedia Commons})
\label{shannon-weaver-model}
%\remarkP{TODO: More neutral skin color.}
}
\end{figure*}

%Anecdotal evidence suggests that running the same test case on
%multiple test systems is a challenge in many organizations.
% At Westermo, we map the network topology of a test case onto that of a
%test system.
\papermapper{} investigates and discusses ``the mapping problem'' and how it is related to the subgraph isomorphism problem.
The subgraph isomorphism problem is a well-studied topic in the field of graph theory, and much of the basic terminology is covered in books such as Wilson's introduction to graph theory~\cite{wilson2010}.
%\remarkD{This is probably not wrong, but I personally dislike when books or articles are only referenced by their citation bracket.}
%\remarkP{done}
%
There are a number of approaches to solving the subgraph isomorphism problem:
satisfiability, for example, is covered by both Knuth
and Ullman~\cite{knuth2015fascicle,ullmann2010bit}.
Another approach is to do a combinatorial search, such
the one described by Bonicci et al.~\cite{bonnici2013subgraph}.
In the test cases and test systems at Westermo, there are
frequently cycles. If there is a cycle in a test case, then
this must be mapped onto a cycle in a test system.
Algorithms for identifying cycles
%, such as the ones described by
%Paton, or Yuster and Zwick,
are therefore of importance for improving the mapping,
e.g.\ work by Paton, or Yuster and Zwick~\cite{paton1969algorithm, yuster1997finding}.

%% The body of knowledge with respect to assigning
%% test environments to test cases for execution
%% seems to be in its infancy, and

\mysubsection{Intermittently Failing Tests}
According to Fowler~\cite{fowler2006codesmell,fowler2018refactoring}, a \textit{code smell} is a ``surface indication'' of a potential problem in code. \textit{Smelly tests} represent code smells in test code, such as test cases with poor design or implementation choices. An example of a test smell is ``the local hero''
test case that will pass in a local environment, but fail when testing in the target environment because of undocumented dependencies that were not satisfied~\cite{garousi2018smells}.
A \textit{flaky test} is a test that has been executed at least twice and provided both passing and failing verdicts, without any changes being made to the underlying software, hardware or testware between the two executions~\cite{luo2014empirical}.
In other words, a flaky test is a test case in which there is an apparent non-determinism in the underlying software, hardware or testware. Flaky tests have been widely studied, and many tests are flaky because they are also smelly~\cite{vahabzadeh2015empirical},
and sometimes the removal of test smells leads to removal of flakiness.

Because of the requirement of unchanged software, hardware and testware, the construct of a flaky test is not very convenient in a resource constrained continuous integration environment.
Therefore, \paperflaky{} defines an \textit{intermittently failing test} as a test case that has been executed repeatedly while there is a potential evolution in software, hardware and/or testware, and where the verdict changes over time.\footnote{A recent paper by Barboni et al.~\cite{barboni2021flakiness} explores different definitions and overlapping aspects of the terminology of flaky or intermittently failing tests.
}
The paper also defines a metric to quantify test cases as more or less intermittently failing based on Markov chains~\cite{markov1906extension, privault2013understanding}.
A similar metric was presented by Breuer in 1973~\cite{breuer1973testing},
but discovered independently, and created for another domain -- \paperflaky{} targets regression testing of software-intense embedded system as opposed to describing faults in hardware.
Breuer seems to have been first to use Markov chains to describe faults in a system, as opposed to modeling the system.
Another way of quantifying flakiness, based on entropy, has been presented by Gao~\cite{gao2017quantifying}. This method requires code instrumentation (code coverage) and was evaluated using Java programs with 9 to 90 thousand lines of code.
One perceived advantage of Gao's metric is that it can be used to ``weed out flaky failures'' whereas our approach targeted a better understanding of the root causes of intermittently failing tests.

Lou et al.~\cite{luo2014empirical} did an investigation on unit tests in open source projects.
They found that many tests were flaky because of asynchronous waiting, concurrency, or test order.
Important problems with flaky tests, on this level, are that they can be hard to reproduce, may waste time, may hide other bugs and they can reduce the confidence in testing such that practitioners ignore failing tests~\cite{luo2014empirical}.
Despite a large amount of research on flaky tests on unit level,
only four studies targeting intermittently failing tests on a system level were identified when 
\paperflaky{} was submitted: %, we could only identify four other papers that target intermittently failing tests on a system level:
Ahmad et al.~\cite{ahmad2021empirical},
Eck et al.~\cite{eck2019understanding},
Lam et al.~\cite{lam2019root}, and
Thorve et al.~\cite{thorve2018empirical}.
These five studies combined allowed for identification of similarities and differences between factors leading to intermittently failing tests on unit and system levels.

\mysection{Communication and Flow of Information}

In a 60 year old publication~\cite{shannon1948} Shannon
first described what would later be known as the Shannon-Weaver model
of communication, shown in Figure~\ref{shannon-weaver-model}.
This model
%is said to be the mother of the area of Information Technology (IT),
%and
has later been built upon to represent
countless variants.
It contains
an information source converted into a message that a
transmitter converts to a signal,
noise is added to the signal from a noise source.
The signal and message then reach the destination.
Later research on this model add,
among other things, medium used,
as well as the role of distances.
%% perhaps as a type of noise between sender and receiver.
%% %
%% For example:
%% Innis mentions that the \emph{medium} used can play a great role
%% and may have different characteristics~\cite{innis2007empire}.
%% %\remarkP{Tom was not fond of this next sentence.}
%% %For ancient empires papyrus could be used to reach over distances,
%% %whereas a pyramid instead may reach over time.

Information flow has an important overlap with the concept of
communication, it can be defined by a distributed system of agents and
the relationships between them. Information flow is important when a
synergy between humans and software systems is required for a work
flow~\cite{durugbo2013modelling}.
%
%% In the daily work flow of tasks relevant to software engineering
%% practitioners, there is a need to make minor decisions such as: ``what
%% should I implement, correct or test next?''  and also major ones such
%% as: ``is the software of good quality, have we tested enough, can we
%% release it?''

%% A modern communication medium
%% frequently relies on technical solutions,
%% such as a conference phone.
%% Olson and Olson
%% mention that one type of
%% \emph{noise}, or failure of remote work
%% is related to bad usage of technology:
%% ``[Users of the conference phone]
%% adapted their behavior rather than fix the technology. On many
%% occasions, the participants shouted because the volume
%% [of the conference phone] was set too low''~\cite{olson2000distance}.
%% We speculate that other communication media,
%% such as an automatically generated test report in PDF format,
%% a test results trend plot rendered in-browser with JavaScript,
%% or a manually written test report all have an impact on
%% communication, and that users of these in many cases
%% have adapted to them, as opposed to improved upon any flaws in them.

A theory for
\emph{distances} in software engineering ``explains how practices improve
the communication within a project by impacting distances between
people, activities and artifacts''~\cite{bjarnason2015role}.
As an example, the practice of
\emph{Cross-Role Collaboration} has an impact on the temporal distance.
Olson and Olson found that
new technology may make geographical distance smaller, making cultural
distances appear to be greater~\cite{olson2000distance}. %: ``Mexican engineers in khaki shirts
%and sunglasses looked suspicious to the shirt-and-tie
%U.S. engineers.''
They also found that modern communication media
frequently relies on technical solutions,
such as a conference phone,
and when there are usability problems,
%Olson and Olson
%mention that one type of
%\emph{noise}, or failure of remote work
%is related to bad usage of technology:
the users adapted their behavior instead of fixing the technology --
by shouting in the conference phone.
%On many
%occasions, the participants shouted because the volume
%[of the conference phone] was set too low''~\cite{olson2000distance}.

The requirements specification is the most important document for testers
during system testing,
but this document is often of poor quality~\cite{illes2008role},
and is one of the main sources of technical debt for safety development~\cite{ghanbari2016seeking}.
Alternatively,
documentation can be seen as a complement to communication that can help combat knowledge evaporation~\cite{manai2019software}.
Human factors and the quality of the requirements specifications
were also mentioned in a qualitative interview study
``as essential for development of safety critical systems''~\cite{notander2013challenges}.
%% In the same paper:
%% ``Development of safety-critical systems is not about writing code. It
%% is about understanding the problem that should be solved by the system
%% and be aware of the special nature of safety-critical systems\ldots''
Furthermore,
the Annex E of the first part of the ISO/IEC/IEEE 29119 standard
acknowledges that testers need to communicate, with various stakeholders,
in a timely manner,
and that this communication may be more formal (e.g., written reports)
or oral (as in agile contexts)~\cite{iso291191}.
Similarly, the International Software Qualifications Board, ISTQB,
released their syllabus
for certification of test automation engineers~\cite{ISTQB-TAE} in 2016.
It covers topics that frequently occurred in our interviews
in \papertrc.
They recommend that one should:
measure benefits of automation (costs are easily seen),
visualize results,
create and store logs from both the system under test
and the test framework, and
generate reports after test sessions.

%% We also note, as pointed out by Garousi and Mäntylä~\cite{garousi2016}, that there is a need for secondary studies in the
%% field of test reporting, as described in a recent
%% Systematic Literature Review (SLR) of SLRs in
%% software testing.

Papers \ptrc{} and \ptim{} extend the body of knowledge
in the field of communication and information flow in software
engineering with results on a holistic
approach to the flow of information in software testing,
as well as a tool for test results exploration and visualization.

\mysection{Visualization}

%% Source code can be hard to grasp,
%% and source code visualization might help.
%% In an influential publication from 1992
%% the source code is shown from a helicopter view as thin colored lines.
%% The colors represent an attribute of the code, such as age~\cite{eick1992seesoft}.

According to Diehl, there are three important aspects of
software visualization:
structure, behavior and evolution~\cite{diehl2014past}.
An early visualization technique to show structure is from 1958
and generated control-flow diagrams because it
``is not practical to expect [everyone]
to be intimately familiar with [all code]''~\cite{scott1958automatic}.
Visualizations may not only show structure but also compare structures.
One survey on software visualization,
present methods for visualizing static aspects with the focus of
source code lines, class metrics, relationships, and architectural
metrics~\cite{caserta2011visualization}.

In a well-cited paper from 2002, Jones et al.\ use test results data with
a combination of source code, code coverage, unit tests and test
results in a sort of an extended Integrated
Development Environment (IDE) view~\cite{jones2002visualization}.
A dashboard is a common approach for quality monitoring and control in
the industry. These aim at presenting one or more key performance
indicators (KPIs) over time~\cite{deissenboeck2008tool,froese2016lessons}.
%%
%% Color blindness is a condition where an individual cannot fully
%% see color or differences in color. It was first described by
%% John Dalton in 1798~\cite{dalton1798}. Roughly 8\% of all men
%% and 0.5\% of all women have some form of color blindness~\cite{kalloniatis2007perception}.
%% %
%% When a difference in colors is used to carry information there
%% is a risk that color blind individuals cannot receive the
%% information or receives distorted information. This is used in
%% the classic Ishihara test from 1917
%% where military examiners became a special target audience~\cite{ishihara1917tests}.
%% Technology is also important to consider when deciding on
%% colors for visualizations:
%% ``different areas of a plot should still be distinguishable when
%% the graphic is displayed on an LCD projector rather than a
%% computer screen, or when it is printed on a grayscale printer,
%% or when the person viewing the graphic is color-blind''~\cite{zeileis2009}.
%%
When visualizing test results,
%
%(i)
there is a need for summaries~\cite{nilsson2014visualizing},
%
%(ii)
information may be spread out in several different systems~\cite{brandtner2014supporting},
%
%(iii)
one might need to consider transitions between different views of a
test results visualization tool~\cite{opmanis2016visualization},
and
%
%(iv)
one should consider that visualizations may target different stakeholders
for different purposes~\cite{strandberg-heatmaps}.
% Many of there studies also propose visualization methods.

%\remarkP{Also cite, Rosling, Rosling and Rosling?
%Rosling, H., Rosling, R. A., and Rosling, O. (2005). New software brings statistics beyond the eye. Statistics, Knowledge and Policy: Key Indicators to Inform Decision Making. Paris, France: OECD Publishing, 522-530.}
%

Rosling et al.\ argues that ``The world cannot be understood without numbers. But the world cannot be understood with numbers alone.''
The Roslings developed software to visualize the improvements in world health because students were found to perform worse than random in quizzes on the topic
\cite{rosling2005new,rosling2018factfulness}.
This could be seen as an indication of the importance of visualizations for comprehending data.
In his keynote at
the practitioner-oriented conference Beauty in Code 2018~\cite{bach2018blink},
James Bach highlighted that
stakeholders do not know what to visualize:
``What I had to do, is use my skill as a tester, and
my interest in visual design, and complexity, and displays,
and statistics, to try to come up with something, that they,
when they saw it -- it would be like someone seeing an iPhone
for the first time, they would say `I used to like Nokias and now I
like Apple phones'.''
This could be seen as supporting findings in \papertrc,
that visualizations are not prioritized and an individual
skilled in other domains ends up being the one preparing visualizations.

Adopting a data science approach at a company can be hard -- data science deals with collection, processing, preparation and analysis of data, it is a cross-domain discipline that requires skills beyond software engineering and computer science, e.g.\ mathematics and domain knowledge~\cite{ebert2019data}.
\papertim{} strives to make test results data visualized and explorable. Similar goals were held in the
q-rapids project\footnote{Q-Rapids: Quality-aware rapid software development: \href{https://www.q-rapids.eu/}{www.q-rapids.eu}.
%\remarkW{add website in footnote?}
%\remarkP{Done.}
}
where data was collected, processed and then shown to product owners with the aim of enhancing their cognition, situational awareness, and decision-making capabilities~\cite{choras2019increasing}. Some of the challenges they have seen in their work on q-rapids are:
to be transparent about what a value or metric means and its origins, integration with the tool and other tools, as well as the need for the tool to be tailored to a company.
Some of the lessons learned were to use a common entry point for the data, and that implementation of the tool requires experts (e.g.\ when setting up data collection streams and doing analysis of data)~\cite{mf2019continuously}.
As opposed to the q-rapids project, the tool described in  \papertim\ does not strive to distill test results data into key performance indicators. Instead, it targets exploration and visualization and, in some sense, interaction with the data.
Ahmad et al.\ \cite{ahmad2021data} investigated information needs for testing and identified eight needs that closely match the work in \papertim, such as being aware of if, and if so where (on which test system and code branch), a test case fails. 

%\remarkP{Mention something cool from the Tim paper here.}

With respect to the body of knowledge in the field of
visualization of software test results, \papertim{}
confirms some of the previous work (such as the importance of visualizations,
and the need to consider transitions between views).
It also provides a long term case study of the implementation and usage
of test results visualization and decision making,
and shows the importance of the TRDB. % in \papertim{}. % and \ptrdb.
%\remarkW{(Paper X2?)}
%\remarkP{Done}

%\remarkP{TODO 20210708: cite \cite{ahmad2021data} here.
%Discuss how strikingly their findings overlap with ours?
%}

%%%%%%%%%%%%%%%%%%%%%%%%%%%%%%%%%%%%%%%%%%%%%%%%%%%%%%%%%%%%%%%%%%%%%%%%%

%\newpage
\mychapter{Contributions}
\label{kappa-contributions}

The research goal of this thesis is
\emph{to improve automated system-level software testing of industrial networked embedded systems.}
To reach the goal, the thesis poses five RQs that have been targeted with five studies. Each paper primarily targets one RQ (RQ1: \papertrc, RQ2: \paperissre, etc.).
This chapter revisits the RQs and summarizes their main contributions (C).
% \\ \remarkP{re-written 2021-09-01 11:24}

\mysection{RQ1: Information Flow}
\emph{How could one describe the flow of information in software testing in an organization
  developing embedded systems, and what
  key aspects,
  challenges,
  and
  good approaches
  are relevant to this flow?}

\papertrc{} seems to be the first study to
take a high level approach to the flow of information in software testing.
For the study, 
twelve interviews with industry practitioners were held.
The interviewees were from five different companies and had an average of more than 14 years of experience.
17 hours of audio was recorded, anonymized and transcribed into 130 pages of text,
that was analyzed with thematic analysis.
The main contributions of \papertrc{} are:

%%%%% 
%%%%% \papertrc{} seems to be the first study to
%%%%% take a high level approach to the flow of information in software testing.
%%%%% %
%%%%% For the study, 
%%%%% twelve interviews with industry practitioners were held.
%%%%% The interviewees were from five different companies and had an average of more than 14 years of experience.
%%%%% 17 hours of audio was recorded, anonymized and transcribed into 130 pages of text,
%%%%% that was analyzed with thematic analysis.
%%%%% %
%%%%% The main contributions of \papertrc{} are:

\begin{description}
\item[C-A1:]{%
An overall model of the flow of information in software testing. % was identified. %, see Figure~\ref{fig-contrib-flow}.
}
\item[C-A2:]{%
Six key factors that affect the flow of information in software testing
%are:
(how organizations conduct testing and trouble shooting,
communication,
processes,
technology,
artifacts, as well as how it is organized).
}
\item[C-A3:]{%
Seven main challenges for the information flow
(comprehending the objectives and details of testing,
root cause identification,
poor feedback,
postponed testing,
poor artifacts and traceability,
poor tools and test infrastructure, and
distances).
}
\item[C-A4:]{%
Five good approaches for enhancing the flow
%were:
(close collaboration between roles,
fast feedback,
custom test report automation,
test results visualization, and
the use of suitable tools and frameworks).
}
\end{description}

%% \remarkP{From text to bullets?}
%% \sout{
%% From the data, an overall model of the flow of information in software testing was identified, see Figure~\ref{fig-contrib-flow}.
%% Six key factors that affect the flow of information in software testing are:
%% how organizations conduct testing and trouble shooting,
%% communication,
%% processes,
%% technology,
%% artifacts, as well as how it is organized.
%% The main challenges for the flow are:
%% comprehending the objectives and details of testing,
%% root cause identification,
%% poor feedback,
%% postponed testing,
%% poor artifacts and traceability,
%% poor tools and test infrastructure, and
%% distances.
%% Finally, five good approaches for enhancing the flow were:
%% close collaboration between roles,
%% fast feedback,
%% custom test report automation,
%% test results visualization, and
%% the use of suitable tools and frameworks.
%% \papertrc{} seems to be the first study to
%% take a high level approach to the flow of information in software testing.
%% }

\mysection{RQ2: Test Selection}
\emph{What challenges might an organization have with respect
    to system-level regression test selection in the context of
    networked embedded systems,
    and how could one address these challenges?
}

\paperissre{} seems to be one of the first papers on system-level regression test selection. The study covers three industrial challenges with nightly regression testing:
nightly testing not finishing on time,
manual work and omitted tests, as well as
no priority for the test cases.
These problems were solved by implementing SuiteBuilder.
The algorithm was evaluated quantitatively using data from four years
of nightly testing, as well as qualitatively with interview data.
The main contributions of \paperissre{} are:

%%%%% 
%%%%% \paperissre{} seems to be one of the first papers on system-level regression test selection. The study covers three industrial challenges with nightly regression testing:
%%%%% nightly testing not finishing on time,
%%%%% manual work and omitted tests, as well as
%%%%% no priority for the test cases.
%%%%% %
%%%%% These problems were solved by implementing SuiteBuilder.
%%%%% The algorithm was evaluated quantitatively using data from four years
%%%%% of nightly testing, as well as qualitatively with interview data.
%An experimental evaluation of the SuiteBuilder tool found that test suites finish on time,
%and that two thirds of the failing tests
%are now positioned in the first third of the test suites.
%%%%% %
%%%%% The main contributions are:
%%%%% \remarkP{TODO: Rephrase these, improve.}

\begin{description}
\item[C-B1:]{%
%A description of t
The SuiteBuilder tool, a framework
of prioritizers assigning priorities based on
multiple factors.
}
\item[C-B2:]{%
Empirical evidence that the tool improves the testing process: test suites finish on time,
and that two thirds of the failing tests
are now positioned in the first third of the test suites.
}
%\item[C-B4:]{%
%Demonstration of feasibility of \emph{system-level} regression test selection.
%}
\end{description}

%% 
%% \sout{
%% \paperissre{} covers
%% % \remarkD{Wouldn't you say that the focus of Paper B was motivated by these three challenges, or that it addressed them? We didn't really \textit{identify} them as part of the work in this particular paper}
%% % \remarkP{I think we twisted the RQ's a bit, so I think the current text is pretty OK.}
%% three industrial challenges with nightly regression testing:
%% nightly testing not finishing on time,
%% manual work and omitted tests, as well as
%% no priority for the test cases.
%% These problems were solved by implementing SuiteBuilder, a framework
%% of prioritizers that assign priorities based on
%% multiple factors.
%% The algorithm was evaluated quantitatively using data from four years
%% of nightly testing, as well as qualitatively with interview data.
%% An experimental evaluation of the SuiteBuilder tool found that test suites finish on time,
%% and that two thirds of the failing tests
%% are now positioned in the first third of the test suites.
%% Another contribution of this work
%% is the \emph{system-level} regression test selection,
%% and an implementation that solves these three problems
%% in our industrial context.
%% %Work could be done on our approach, perhaps additions or removals of
%% %prioritizers could be suitable to other environments,
%% %depending on available data.
%% }

\mysection{RQ3: Hardware Selection}
\emph{What challenges might an organization have with respect
    to test environment assignment in the context of
    networked embedded systems,
    and how could one address these challenges?}

%%%%%%% The process of mapping involves how one maps the needs of a test case onto a test system. \papermapper{} explores this problem 
%%%%%%% by using graph theory, in particular the graph models of the test systems and test cases. This way the subgraph isomorphism problem is adapted to the context of networked embedded devices.
%%%%%%% A prototype implementation was evaluated quantitatively with the
%%%%%%% available test systems and test cases for more than 10.000 different pairs of graphs. 
%%%%%%% The main contributions of \papermapper{} are:

%%%%% 
%%%%% %\remarkP{Hmmm, I'm not sure this text matches the RQ any more.}
When performing testing of embedded systems, it is critical
to also involve real hardware, and \papermapper{}
shows what appears to be the first evaluated approach to solve the mapping problem. The process of mapping involves a way to map a test case onto a test system using graph theory, in particular the graph models of the test systems and test cases. This way the subgraph isomorphism problem is adapted to the context of networked embedded devices.
The prototype implementation was evaluated quantitatively with the
available test systems and test cases for more than 10.000 different pairs of graphs. 
The main contributions of \papermapper{} are:
%%%%% The main contributions are:
%%%%% 

\begin{description}
\item[C-C1:]{%
A new mapper, that was found to be more than 80 times faster than the old tool.
}
\item[C-C2:]{%
An extension of the mapper, where the TRDB with previous mappings is used to
%By using a TRDB with previous mappings, the algorithm could 
map in
\emph{different} ways over time such that %. This way 
the DUT coverage grew from
a median of 33\%, to a median of 100\% in just five iterations.
}
\end{description}

%% 
%% \sout{There seems to be no prior evaluated solution in the domain of networked embedded devices.
%% The approach described uses the graph models of the test systems
%% and test cases, and adapt the subgraph isomorphism problem to the context of networked embedded devices.
%% The prototype implementation quantitatively was evaluated with the
%% available test systems and test cases for more than 10.000 different pairs
%% of graphs. The new tool was found to be more than 80 times faster than the old one.
%% Also, by using a TRDB with previous mappings, the algorithm could map in
%% different ways over time. This way the DUT coverage grew from
%% a median of 33\% for the 10.000 pairs, to a median of 100\%
%% in just five iterations.
%% }

\mysection{RQ4: Intermittently Failing Tests}
\emph{What are the root causes of intermittently failing tests during system-level testing in a context under evolution, and could one automate the
detection of these test cases?
% What are the root causes of intermittently failing tests during system-level testing
% in a context where the environment is under evolution,
% and could one automate the detection of these test cases?
}

\paperflaky{} defines a novel metric for intermittently failing tests.
By analyzing more than half a million test verdicts from nine months of nightly testing, both intermittently and consistently failing tests were identified. The main contributions of \paperflaky{} are:

\begin{description}
\item[C-D1:]{%
A novel metric that can identify intermittently failing tests, and measure the level of intermittence over the test base.
}
\item[C-D2:]{%
Nine factors that may lead to intermittently failing tests for
  system-level testing of an embedded system
(test case assumptions,
  complexity of testing,
  software and/or hardware faults,
  test case dependencies,
  resource leaks,
  network issues,
  random numbers issues,
  test system issues, and
  refactoring).
}
\item[C-D3:]{%
Evidence that finding the root cause of intermittently failing tests often involves more effort, 
when compared to root cause analysis of consistently failing tests.}
\item[C-D4:]{%
Evidence that a fix for a consistently failing test
  often repairs a larger number of failures found by other test cases
  than a fix of an intermittently failing test.
}
\end{description}

%% \sout{
%% The root causes were analyzed and the paper identifies
%% % factors
%%   nine factors that may lead to intermittently failing tests for
%%   system-level testing of an embedded system:
%%   test case assumptions,
%%   complexity of testing,
%%   software and/or hardware faults,
%%   test case dependencies,
%%   resource leaks,
%%   network issues,
%%   random numbers issues,
%%   test system issues, and
%%   refactoring.
%% \paperflaky{} also identified that
%% finding the root cause of intermittently failing tests often involves more effort, 
%% when compared to root cause analysis of consistently failing tests. This may
%%   lead testers to give up without identifying the failure's root
%%   cause, and to rely on the hope that failures will occur infrequently
%%   enough that they can be ignored.
%% Finally, the paper identifies that a fix for a consistently failing test
%%   often repairs a larger number of failures found by other test cases
%%   than a fix of an intermittently failing test.
%% }

\mysection{RQ5: Test Results Exploration and Visualization}

\emph{How could one implement and evaluate a system to enhance
  visualization and exploration of test results to support the information
  flow in an organization?}

%A TRDB plays a central role in the testing of embedded systems,
%both for its use in regression test selection,
%and also for the flow of information in software testing.
\papertim{} focuses on a new tool, Tim, that was implemented to replace an old system for test results exploration and visualization (TREV).
Work was prioritized and evaluated with a reference group of 12 individuals.
From reference group meetings, 7 hours of video was recorded and transcribed into 36 pages of text,
and logs from 201 days of using Tim was also collected.
The main contributions of \papertim{} are:

\begin{description}
\item[C-E1:]{%
%The identification of f
Four solution patterns for TREV (filtering, aggregation, previews and comparisons).
}
\item[C-E2:]{%
%The i
Implementation and empirical evaluation of eight views for TREV (start, outcomes, outcome, session, heatmap, measurements, compare branch and analyze branch views).
}
\item[C-E3:]{%
%The identification of s
Six challenges for TREV (expectations, anomalies, navigation, integrations, hardware details and plots).
}
\end{description}

\mychapter{Future Work}

% \remarkD{***** DANIEL HERE IN REVIEW *****}

This thesis covers system-level test automation of embedded systems, and in particular test results information flow, regression test selection, hardware selection, intermittently failing tests as well as test results exploration and visualization. This chapter discusses some of the potential for future research. % in these topics, as well as some others.

The topics of test results information flow and test results exploration and visualization are partly overlapping. Future work could investigate, on a higher level, how an organization would benefit from automated test reporting, as well as if or how humans and such a system could co-create meaningful test reports -- would this be a way to bridge some of the challenges of combining agile and traditional development?

Despite already being very well-researched, there is room for more work on the regression test selection problem -- in particular with respect to diversity, what the characteristics of a good selection is, and how different selection strategies impact these characteristics. Perhaps industry and academia could co-create an open source tool that could be adjusted to fit different needs (such as when one organization requires a certain code coverage, whereas another focuses on diversity)? Also, if the test selection strategy was reactive (perhaps search-based) and given the mandate to re-prioritize on the fly during a test session, would that improve test coverage, or other desirables of the testing?

As mentioned above, research on automated hardware selection for testing embedded systems seems to be in its infancy.
However, perhaps a more generalized problem of hardware selection could be well served from findings in combinatorial testing of product lines or search-based software engineering 
\cite{do2014strategies, hagar2015introducing, harmanjones2001search, lopez2015systematic}?
%
%\remarkD{Well, yes, but it is quite related to combinatiorial testing of product lines.}
%\remarkP{I feel that this is not a simple problem. If we add combinatorial testing and product lines, then why not also search-based software engineering  \cite{do2014strategies, hagar2015introducing, harmanjones2001search, lopez2015systematic}?}
%
With respect to the results in this thesis, questions that remain unanswered range from fundamental (is test coverage of two different hardware platforms exactly twice as valuable as coverage of one?), via detailed (how different would a tool using satisfiability to solve the mapping problem be when compared to one that uses the subgraph isomorphism problem, and are they equally explicable to practitioners?), to extremely detailed (how much does the performance of the mapping tool increase if cycles in the test systems and test cases are considered?).
The benefits and drawbacks of testing on actual devices with hardware, on virtualized systems, or using emulated hardware would also be interesting to explore. When is it good enough to test without real hardware, and when must it be used?

Similar to the regression test selection problem, the topic of intermittently failing tests is receiving a great amount of research attention. Most of this research is done for unit-level testing, perhaps because most test automation is done at this level and public data sets are available. Some of this research seems to imply that the root cause of the intermittence is in the test cases themselves, which \paperflaky{} identified as only one type of root cause. A question that seems to remain unanswered is: which of the root causes to intermittence are generalizable across testing levels and domains?

Rule-based systems, such as the SuiteBuilder tool, is a type of artificial intelligence (AI). More advanced AI learns from data to become increasingly better at solving tasks, sometimes at the price of transparency or explicability \cite{legghutter2007collection,strandberg2021ethical}. It is very clear that society in general, and software test automation in particular, could gain tremendously by adopting more AI. As this thesis is being written, Westermo has joined an industry-academia research project involving AI (described in \cite{eramo2021aidoart}). What the exact outcomes will be remains to be seen, but in general, it seems as if almost all aspects of this thesis could benefit from further automation, with or without AI. 

%\newpage
\mychapter{Conclusions}

As mentioned in the abstract of this thesis, embedded systems appear almost everywhere and play a crucial role for industrial and transport applications -- software testing is crucial for the quality assurance of these systems. This thesis addresses five challenges for automated system-level software testing of industrial networked embedded systems. These challenges have been tackled in five studies, each self-contained with its own conclusion section. 
To conclude the thesis, this chapter takes a step back, and looks at the big picture. 
%return to the overall research goal and discuss unanswered questions suitable for future work.
%
%%%%%%%%%%%%%%%%%%%%%%%%%%%%%%%%%%%%%%%%%%%%%%%%%%%%%%%%%%%%%%%%%%%%%%%%%
%\newpage
%\subsection{Summary and Conclusions}
%\subsection{Conclusion}
%\section{Summary}
%\remarkD{Maybe this section is not needed for a proposal?}
%\remarkP{Let's keep it.}
%
%
%The big picture, 
%
A central result of this thesis is that software testing can be seen as a set of feedback loops,
that can be visualized in the flow diagram
in Figure~\ref{fig-contrib-flow}, which was one of the main findings of \papertrc.
% This figure ties the papers of this thesis together, is visualized by the
% main finding of \papertrc.
This study focused on the overall flow of information in software testing.
The remaining studies fit well into this picture.
Both test selection (\paperissre) and test
environment assignment (\papermapper) can be seen as processes having an impact
on \emph{what} to test, and \emph{with what} to test, in the test environment.
The role of the test results database has, in one way,
an importance \emph{after} the testing (e.g.\ to aid in reporting),
but because it may impact coming test selection and the coming test environment
assignment, it also has an importance \emph{before} the testing activity.
The impact of intermittently failing tests (\paperflaky) can be seen as having an influence
on test results, how it reaches developers and the distrust of testing it
might lead to for developers.
The findings on intermittently failing tests also point to a need to improve software, hardware and testware
in order to avoid intermittent faults in these.
The work on visualizing test results and making them explorable (\papertim)
is one approach of improving the feedback loop in Figure~\ref{fig-contrib-flow}.

The research goal of this thesis is:
\emph{to improve automated system-level software testing of industrial networked embedded systems.}
In short, this has been achieved with an improved understanding of the flow of information in software testing, selection of test cases and hardware for testing, with an investigation of intermittently failing tests, as well as improved visualizations of test results.

\section*{Acknowledgments}
My research was funded by
Westermo %(first under the name of Westermo Research and Development AB, and later as Westermo Network Technologies AB)
and the Swedish Knowledge Foundation through grants
20150277 (ITS ESS-H), and 20160139 (TESTMINE).

\begin{small}
\bibliographystyle{abbrv}
\bibliography{my-references.bib}
\end{small}

\end{document}